\documentclass[12pt]{article}
\usepackage{latexsym, amsmath, epsfig}
\DeclareFontFamily{OT1}{rsfs10}{}
\DeclareFontShape{OT1}{rsfs10}{m}{n}{ <-> rsfs10 }{}
\DeclareMathAlphabet{\mathscript}{OT1}{rsfs10}{m}{n}

\if@twoside\oddsidemargin25mm
\else\oddsidemargin25mm\evensidemargin25mm\marginparwidth25mm\fi
\def\bpl{\Big(}
\def\bpr{\Big)}

\def\t{\theta}

\def\brr{\begin{equation}}
\def\err{\end{equation}}
\def\brr{\begin{eqnarray}}
\def\err{\end{eqnarray}}
\def\ba{\left(\begin{array}}
\def\ea{\end{array}\right)}

\newcommand{\dr}{\raise.3ex\hbox{$\stackrel{\leftarrow}{\partial }$}{}}
\newcommand{\dl}{\raise.3ex\hbox{$\stackrel{\rightarrow}{\partial}$}{}}

\newcommand{\ft}[2]{{\textstyle\frac{#1}{#2}}}
\newcommand{\ns}{\normalsize}

\renewcommand{\a}{\alpha}
\renewcommand{\b}{\beta}

\renewcommand{\d}{\delta}

\newcommand{\s}{\sigma}

\newcommand{\redx}{\includegraphics[width=.15in,angle=0]{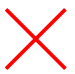}}

\begin{document}


\begin{titlepage}

\vspace{-3cm}

\title{
   \hfill{\ns CU-TP-977\\}
   \hfill{\ns HU-EP 00/23\\} 
   \hfill{\ns UPR-889T\\} 
   \hfill{\ns hep-th/0005251\\[1cm]}
   {\LARGE  Local Anomaly Cancellation, {\it M}-Theory Orbifolds and 
            Phase-Transitions \\[.5cm]  }}
                       
\author{{\bf
   Michael Faux$^{1}$, \,
   Dieter L{\"u}st$^{2}$ \,
   and Burt A.~Ovrut$^{3}$}\\[5mm]
   {\it $^1$Departments of Mathematics and Physics} \\
   {\it Columbia University} \\
   {\it 2990 Broadway, New York, NY 10027} \\[3mm]
   {\it $^2$Institut f\"ur Physik, Humboldt Universit\"at} \\
   {\it Invalidenstra\ss{}e 110, 10115 Berlin, Germany} \\[3mm]
   {\it $^3$Department of Physics, University of Pennsylvania} \\
   {\it Philadelphia, PA 19104--6396, USA}}
\date{}

\maketitle

\begin{abstract}
\noindent
In this paper we consider orbifold compactifications of M-theory on 
$S^1/{\bf Z}_2\times T^4/{\bf Z}_2$. 
We discuss solutions
of the local anomaly matching conditions
by twisted
vector, tensor and hypermultiplets 
confined on the local orbifold six-planes.  
In addition we consider phase-transitions among different solutions which are
mediated by M-theory fivebranes which touch the local orbifold planes
and are converted there to gauge instantons. 
\end{abstract}

\thispagestyle{empty}

\end{titlepage}


\section{Introduction}

\noindent
{\it M}-theory harbors a broad spectrum of phenomena which can
be systematically probed by analyzing anomalies in effective 
quantum field theories.  In the case of orbifold compactifications
of eleven-dimensional supergravity, a wide range of topological 
restrictions can be resolved, and the states localized
on orbifold planes determined, by imposing factorization criteria 
on anomaly polynomials.
The basic paradigm was espoused in \cite{hw1, hw2} by analyzing the
$S^1/{\bf Z}_2$ compactification. This was successfully applied to 
the $T^5/{\bf Z}_2$ compactification in \cite{dasmuk, wittens5}.   
In each of these cases, the orbifold planes comprise
isolated, non-intersecting submanifolds.  
In more general situations, the orbifold planes can 
intersect, which gives rise to a number of novel features.
In this paper, we describe local anomaly cancellation on
$S^1/{\bf Z}_2\times T^4/{\bf Z}_M$ compactifications of
{\it M}-theory, with $M=2,3,4$ or $6$. These correspond to
special points in the moduli space of $S^1/{\bf Z}_2\times K3$
compactifications.  In such situations, the orbifold planes 
do intersect. These issues were first discussed in \cite{mlo} and later
in \cite{ksty,mo}. Here, we greatly extend the analysis 
of local anomaly cancellation in orbifolds of this type, particularly 
emphasizing results pertaining to the simplest of these cases, 
corresponding to $M=2$.

In our previous work \cite{mlo}, we described the general features of
$S^{1}/{\bf Z \rm}_{2} \times T^{4}/{\bf Z \rm}_{M}$ ${\it M}$-theory orbifolds.
In this paper, we expand those results, incorporating
two subtle technical points which were not addressed in complete generality
in our previous work.  The first is an issue pertaining to the periodicity of 
the four-form $G$ that has recently been more thoroughly described in 
\cite{bds}.  The second issue concerns the precise normalization 
of the $CGG$ term in the supergravity action.  Each of these 
impinge numerically on our analyses, both in \cite{mlo} 
and in this current paper, 
by changing the overall coefficient of the anomaly inflow due
to the classical variation of the $CGG$ term. We treat this
coefficient as a parameter, to be determined by consistency
arguments, in a manner similar to the approach described in \cite{wittenflux}.
In order that ${\it M}$-fivebranes have unit magnetic charge,
we choose a scale for the three-form potential $C$ such that, 
upon integration over dimensions transverse to a fivebrane, we obtain 
$\int dG=1$. This leaves only one coefficient in the
basic Chern-Simons interactions of the effective field theory
unconstrained by supersymmetry and by the requirement of fivebrane anomaly 
cancellation. However, this one parameter is uniquely fixed by the additional 
requirement of consistent orbifold compactifications and gives rise to the 
particular coefficient cited below for the $CGG$ inflow anomaly.
As a result of these two described 
changes we now find a whole class of solutions
of the local anomaly matching conditions, which in general consist of
vector, hyper  and tensor multiplets 
confined on the local orbifold six-planes and which also,
for global consistency requirements, contain fivebranes, free to 
move in the 11-dimensional
bulk (see also ref.\cite{ksty} for discussion of solutions of the
local anomaly equations).

This situation is interesting since it is the simplest 
scenario which involves fivebrane-mediated phase transitions
in which the gauge group is nontrivially influenced by local tensor couplings.
Specifically, if a fivebrane 
hits one of the local orbifold six-planes it will be
described by a torsion free sheaf being equivalent to a small gauge instanton.
Due to the presence of this  instanton the original gauge group at the local
orbifold plane will be broken to some subgroup, where also
the number of
tensor and hypermultiplets gets changed. 
In this way elements of a certain class of orbifold
compactifications are
related to each other by fivebrane-mediated phase transitions, where
the associated magnetic charges at the six-dimensional orbifold planes
are changing by one unit. In addition we
also discuss the possibility of phase transitions with half-integer change
of magnetic charge at the local six-planes. We suggest that these
phase transitions are due to fivebranes which split at interconnecting
ten-planes and are then transmuted to half-integrally charged gauge
instantons.

The orbifold compactifications 
which we discuss in this paper
are of interest for other reasons as well.
For example, in a series of papers, the $S^{1}/{\bf Z \rm}_{2}$ orbifold was
compactified on smooth Calabi-Yau threefolds producing 
realistic ``brane universe'' theories of particle physics and cosmology
\cite{bopw}. In future work, we will explore both the formal and
phenomenological aspects of different 
${\it M}$-theory orbifold compactifications.

\section{$S^{1}/{\bf Z \rm}_{2} \times T^{4}/{\bf Z \rm}_{M}$ Orbifolds}

The $S^1/{\bf Z}_2\times T^4/{\bf Z}_M$ orbifolds each involve
a pair of ten-dimensional hyperplanes fixed under the 
${\bf Z}_2$ projection, which we denote by $\a$, 
and a set of distinct seven-dimensional
hyperplanes fixed under the ${\bf Z}_M$ projection, which we denote 
by $\b$.  Each of the $\b$-planes transversally intersects each of 
the $\a$-planes once, at particular six-dimensional hyperplanes invariant
under both $\a$ and $\b$.  Chiral anomalies are induced on the
$\a$-planes and (separately) on the $\a\b$-planes, 
due to localized chiral projections
of fields.  Cancellation of the ten-dimensional $\a$-plane anomalies
is uniquely accomplished by an additional ten-dimensional $E_8$ Yang-Mills
supermultiplet on each of the two $\a$-planes, as is well-known.  In this
paper, we concern ourselves with the six-dimensional $\a\b$-plane 
anomalies.

Although some of our discussion will be more general, it is helpful to 
have a specific orbifold in mind to help visualize the basic geometric
setting.  Our prototype is the simplest of the orbifolds described above,
namely, the $S^1/{\bf Z}_2\times T^4/{\bf Z}_2$ 
orbifold.  In this case, spacetime has topology ${\bf R}^6\times T^5$ and
each of the five compact coordinates  
takes values on the interval $[-\pi,\pi]$ with the endpoints
identified. The nontrivial projections
are $\a: (\,x^\mu\,,\,x^i\,,\,x^{11}\,)\longrightarrow (\,x^\mu\,,\,x^i\,,\,-x^{11}\,)$
and $\b: (\,x^\mu\,,\,x^i\,,\,x^{11}\,)\to (\,x^\mu\,,\,-x^i\,,\,x^{11}\,)$,
where $x^\mu$ parameterizes the six noncompact dimensions,
while $x^i$ and $x^{11}$ parameterize the $T^4$ and $S^1$ factors respectively.
The element $\a$ leaves invariant the two ten-planes
defined by $x^{11}=0$ and $x^{11}=\pi$, while
$\b$ leaves invariant the sixteen seven-planes defined 
when the four coordinates $x^i$ individually assume 
the values $0$ or $\pi$.  Finally, $\a\b$ leaves invariant 
the thirty-two six-planes defined when
all five compact coordinates individually assume the values $0$ or
$\pi$.  The $\a\b$ six-planes coincide with 
intersections of the $\a$ ten-planes with the 
$\b$ seven-planes. The global structure is depicted in Figure 1. 
\begin{figure}
\begin{center}
\includegraphics[width=5in,angle=0]{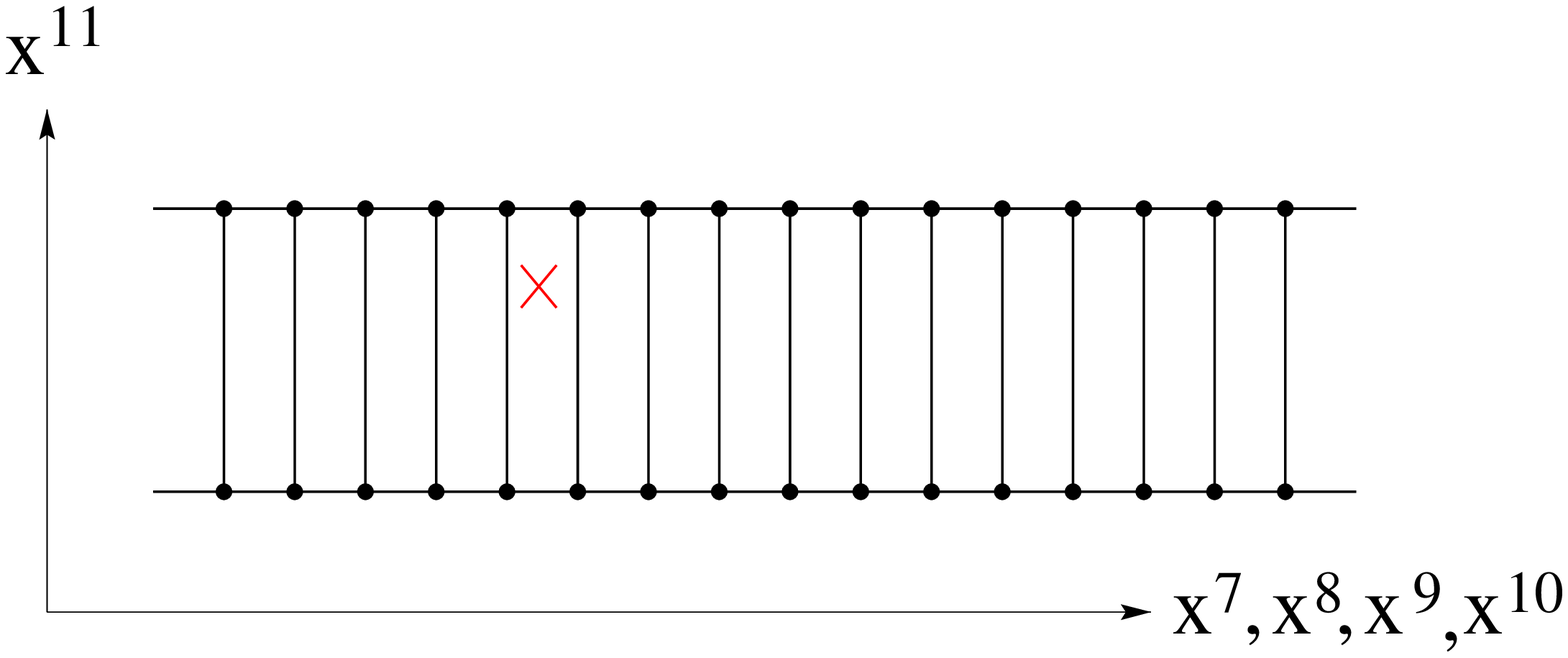}\\[.2in]
\parbox{6in}{Figure 1:  The global structure of orbifold planes
in the $S^1/{\bf Z}_2\times T^4/{\bf Z}_2$ orbifold.  
Horizontal lines represent the two ten-dimensional $\a$-planes, while
vertical lines represent the sixteen seven-dimensional
$\b$-planes. The thirty-two six-dimensional $\a\b$-planes 
are represented by the solid dots and coincide
with the intersection of the $\a$-planes and the $\b$-planes.
The \includegraphics[width=.15in,angle=0]{x.eps} 
in the figure indicates a ``wandering" fivebrane.}
\end{center}
\end{figure}

There are several magnetic and electric sources for $G$ necessary
to resolve chiral anomalies in these orbifolds. The basic 
Chern-Simons terms include the $CGG$ interaction 
and also the higher-derivative $GX_7$ interaction, 
where $X_8\equiv dX_7$ is the
eight-form describing the worldvolume anomaly generated
by the fivebrane zero modes.  The worldvolume anomaly is cancelled by inflow 
mediated by the $GX_7$ interaction, provided the fivebrane 
acts as a magnetic source for $G$, in the sense described above.  
Next, each of the two $\a$-planes supports a ten-dimensional $E_8$
super-Yang-Mills multiplet. They also
provide magnetic sources for $G$ due to the presence of 
terms $\d^{(1)} I_4$ in the $dG$ bianchi identity, where
$\d^{(1)}$ is a one-form brane-current localized on the $\a$-plane 
and $I_4$ is a four-form polynomial 
involving the Lorentz-valued curvature $R$ 
and the local $E_8$ field-strength $F$.  

The seven-dimensional $\b$-planes provide
{\it electric} sources for $G$ via Chern-Simons 
interactions $\int \d^{(4)}\,G\,Y^0_3$, 
where $\d^{(4)}$ is a four-form brane-current localized on the 
seven-plane and $Y_4=d Y^0_3$ is a 
gauge-invariant four-form polynomial.  This polynomial 
involves the curvature $R$ 
and also a field strength ${\cal F}$ associated with 
{\it additional} adjoint super-gauge fields
localized on the seven-plane (with the gauge group determined 
by anomaly cancellation in a manner which we will describe).
This coupling gives rise
to an ``I-brane" effect via interplay with the ten-dimensional
magnetic source (involving $I_4$). This contributes additional inflow 
localized on the six-dimensional intersection of the ten-dimensional
$\a$-plane and the seven-dimensional $\b$-plane
\footnote{See Section 5 of \cite{mlo} for a description of this 
effect}.

The magnetic and electric sources described so far
are encapsulated by the following three polynomials
\brr X_8(R) &=&
    \frac{1}{(2\pi)^3\,4!}\,\bpl\,
    \ft18{\rm tr}\,R^4-\ft{1}{32}({\rm tr}\,R^2)^2\,\bpr 
    \nonumber\\[.1in]
    I_{4}(R,F) &=&
    \frac{1}{16\pi^2}\,\bpl -\ft12\,{\rm tr}\,R^2
    +{\rm tr}\,F^2\,\bpr 
    \nonumber\\[.1in]
    Y_4(R,{\cal F}) &=&
    \frac{1}{4\pi}\,\bpl -\ft{1}{32}\,\eta\,{\rm tr}R^2
    +\rho\,{\rm tr}{\cal F}^2\,\bpr \,.
\label{polys}\err
The precise forms of $X_8$ and $I_4$ are fixed by fivebrane
consistency and ten-dimensional anomaly cancellation,
respectively.  The polynomial $Y_4$
is parameterized by two rational coefficients $\eta$ and $\rho$. These are
determined by further requirements described below. 

Finally, the six-dimensional $\a\b$-planes carry a magnetic charge. This
appears in the $dG$ Bianchi identity as a term $g\,\d^{(5)}$, 
where $\d^{(5)}$ is a five-form
brane-current localized on the $\a\b$-plane and $g$ 
is a rational magnetic charge subject to a quantization condition.
For the case $M=2$, the magnetic charge $g$ should be quarter-integer,
as explained in \cite{mlo}.
The Bianchi identity for the four-form field strength $G$ is,
therefore, given by
\brr dG=\sum_{i=1}^2 I_{4(i)}\,\d^{(1)}_{M_i^{10}}
     +\sum_{i=1}^{2f} g_i\,\d^{(5)}_{M_i^6}
     +\sum_{i=1}^{N_5} \d^{(5)}_{W_i^6} \,,
\label{bianchi}\err
where we have included all of the magnetic sources described
above.  The manifold $M_i^{10}$ is the $i$th $\a$-plane, 
while $M_i^6$ is the $i$th $\a\b$-plane and $W_i^6$ is the worldvolume
of the $i$th fivebrane
\footnote{The pervasive use of the label $i$ 
is merely convenient, and does not
imply any specific correlation between these manifolds.}. 
In most of this paper, our expressions apply to a particular 
$\a\b$-plane. Hence, the label $i$ is implicit but omitted.

The projection $\b$ can independently break the ten-dimensional $E_8$ gauge 
groups on the $\a\b$-planes to maximal subgroups.  Since a ten-dimensional
vector supermultiplet decomposes into one six-dimensional $N$=1 vector
and one six-dimensional hypermultiplet, the breaking pattern
will be characterized by an integer number $V_B$ of 
vector multiplets
and another integer number $H_B$ of hypermultiplets, 
each transforming according to some representation ${\cal R}$ 
of the residual maximal
subgroup of $E_8$.  The projection $\b$ necessarily removes
{\it half} of the $E_8$ degrees of freedom.
But the identity of {\it which} half depends on how $\b$ acts on
the $E_8$ root lattice. 

Chiral projection of the supergravity fields results in
further contributions to the local $\a\b$ anomalies.
These derive from ``untwisted" fields comprising
one universal $N$=1 tensor multiplet and some number $h$ of hypermultiplets.
The value of $h$ depends on which $Z_M$ orbifold is being considered.
For the cases $M$=2 and $M$=3, we have $h$=4 and $h$=2 respectively.
Furthermore, since the local anomaly due to the 
supergravity and residual $E_8$ fields 
arises from the coupling of
fields which are not themselves localized on the $\a\b$-planes,
it involve fractions which reflect the multiplicies 
of the fixed planes. We parameterize this by another integer $f$
corresponding to the number of $\b$-planes associated with the
orbifold in question.  For the cases $M$=2 and $M$=3, we have
$f$=16 and $f$=9 respectively. These correspond to the respective sixteen 
and nine fixed-planes in the $Z_2$ and $Z_3$ orbifold limits 
of the $K3$ manifold
\footnote{The $Z_4$ and $Z_6$ orbifolds involve additional subtlety
which we will not discuss in this paper.}.

Finally, we allow for as yet unspecified $N$=1 supermatter localized on each
$\a\b$-plane.  We call this matter ``twisted", since it is
analogous to twisted sector matter in superstring orbifolds. 
This matter assembles into $n_T$ 
tensor multiplets, $n_V$ vector multiplets and $n_H$ hypermultiplets,
and involves an as yet undetermined ``twisted" gauge group $\tilde{{\cal G}}$.
The vector multiplets transform in the adjoint representation, while
the hypermultiplets transform in an unspecified 
representation $\tilde{\cal{R}}$.

We focus on a particular $\a\b$-plane, and assemble each of the 
contributions to the six-dimensional anomaly localized
on this plane. There are three classical (inflow) contributions, 
described by the polynomials
\brr I_8(CGG) &=& 
     -\pi\,g\,I_4(R,F)^2
     \nonumber\\[.1in]
     I_8(GX_7) &=&
     -g\,X_8(R)
     \nonumber\\[.1in]
     I_8(IB) &=&
     -I_4(R,F)\wedge Y_4(R,{\cal F}) \,.
\label{infs}\err
The first two arise from the variation of the $CGG$ and $GX_7$ terms.
The third arises from the variation of $\int \d^{(4)} G Y^0_3$
and describes the ``I-brane" anomaly
\footnote{The necessity for the ``I-brane" contribution in an orbifold
context was first recognized in \cite{mlo}, and was inspired by
an analogous effect on intersecting D-branes, introduced in
\cite{ghm}.}.
There are also three quantum contributions
\brr I_8(SG) &=& \frac{1}{2 f}\,\bpl\,I_{GRAV}^{(3/2)}\,(R)
     -(1+h)\,I_{GRAV}^{(1/2)}\,(R)\,\bpr  
     \nonumber\\[.1in]
     I_8({\cal G}) &=& \frac{1}{f}\,\bpl\,(V_B-H_B)\,I_{GRAV}^{(1/2)}(R)
     +I_{MIXED}^{(1/2)}\,(R,F)_{\cal R}
     +I_{GAUGE}^{(1/2)}\,(F)_{\cal R}\,\bpr 
     \nonumber\\[.1in]
     I_8({\tilde{\cal G}}) &=& (n_V-n_H-n_T)\,I_{GRAV}^{(1/2)}(R)
     -n_T\,I_{GRAV}^{({\rm 3-form})}(R) \nonumber\\[.1in]
     & & +I_{MIXED}^{(1/2)}(R,{\cal F})_{\tilde{\cal R}}
     +I_{GAUGE}^{(1/2)}({\cal F})_{\tilde{\cal R}}\,.
\label{conts}\err  
The factors $I_{\rm GRAV}$, $I_{MIXED}$ and
$I_{GAUGE}$ which appear in the quantum anomalies
describe one-loop gravitational, mixed
and pure-gauge anomalies. They are attributable to the type of chiral fields
with spin indicated by the superscripts.  
These are determined by index theorems and 
are listed explicitly in appendix C of \cite{mlo}.
The anomaly $I_8({\cal G})$ describes the local anomaly involving whatever
subgroup ${\cal G}\subset E_8$ is left unbroken by $\b$ on 
the relevant $\a\b$-plane.
The subscript ${\cal R}$ indicates the representation content of
the $E_8$ residual subgroup. Hence, the traces over the gauge factors in 
$I_{MIXED}^{(1/2)}(R,F)_{\cal R}$ and 
$I_{GAUGE}^{(1/2)}(F)_{\cal R}$ are traces over 
${\cal R}$.  Similar comments apply to the $\a\b$ twisted sector
with gauge group ${\tilde{\cal G}}$, where the representation content
is indicated by the subscript ${\tilde{\cal R}}$.
\footnote{The technical aspects involved in determining (\ref{infs}) and 
(\ref{conts}) are described in \cite{mlo}.
The determination of $I_8(CGG)$ is more subtle
than indicated in that paper, however, for reasons mentioned previously.
See \cite{bds} for a more thorough description of this one term.
Note that the twisted gauge
group ${\tilde{\cal G}}$ can include factors which coincide with some 
factors in the $E_8$ residual group ${\cal G}$.}

There are numerous unspecified parameters involved in the six 
contributions in (\ref{infs}) and (\ref{conts}).  
To begin with, there are ten integers characterizing the global 
geometry of the orbifold,
the local $E_8$ breaking pattern and multiplicities
in the twisted and untwisted spectra. 
For instance, the orbifold geometry is partially
encoded in the number $f$ of $\b$-fixed planes and the number $h$ of
universal untwisted hypermultiplets.
The magnetic charge $g$ is an integer times
a basic quantization unit (which also reflects the orbifold
geometry). 
The two parameters $\eta$ and $\rho$ 
describe electrical charges of the seven-planes, as defined in equation
(\ref{polys}).
The $E_8$ breaking pattern is encoded in
the multiplicities $V_B$ and $H_B$.  Finally, the local twisted spectrum is
encoded in the multiplicities $n_V, n_H$ and $n_T$.
\begin{figure} 
\begin{center}
\begin{tabular}{cc|ccc}
${\cal G}$ & ${\cal R}$ & 
$I_4({\cal R})$ & $I_{2,2}({\cal R})$ & $I_2({\cal R})$ \\[.1in]
\hline
$E_8$ & {\bf 248} & 0 & 9 & 30 \\[.1in]
$E_7$ & {\bf 133} & 0 & 1/6 & 3 \\[.1in]
      & {\bf 56}  & 0 & 1/24 & 1 \\[.1in]
$SO(16)$ & {\bf 120} & 8 & 3 & 14 \\[.1in]
& {\bf 128} & -8 & 6 & 16 \\[.1in]
& {\bf 16} & 1 & 0 & 1 \\[.1in]
$SU(2)$ & {\bf 3} & 0 & 8 & 4 \\[.1in]
& {\bf 2} & 0 & 1/2 & 1 \\[.1in]
\end{tabular}\\[.2in]
\parbox{3.5in}{Table 1: Some useful representation indices.}
\end{center}
\end{figure}

Furthermore, there are six sets of rational parameters which 
characterize the representation content of the matter fields.  These are given by 
``representation indices", which allow one to relate traces over a given
representation to traces over the fundamental representation.
The relevant representation indices are
denoted ${\cal I}_2({\cal R})$,
${\cal I}_{2,2}({\cal R})$ and ${\cal I}_4({\cal R})$,
and are defined by 
\brr {\rm tr}_{\cal R}\,F^2 &=&
     {\cal I}_2({\cal R})\,{\rm tr} F^2
     \nonumber\\[.1in]
     {\rm tr}_{\cal R}\,F^4 &=&
     {\cal I}_{2,2}({\cal R})\,({\rm tr} F^2\,)^2
     +{\cal I}_4({\cal R})\,{\rm tr}F^4 \,,
\err
where all traces on the right-hand side are over the fundamental representation.
Some useful representation indices are listed in Table 1
\footnote{The evaluation of the representation indices
for arbitrary representations of the classical Lie groups
is a complicated problem.  See   
\cite{rsv} for a discussion of this issue.}.

Remarkably, each and every one of the above parameters can be
resolved, in a sense to be made clear, by imposing necessary 
factorization properties on
the net anomaly obtained by summing the six contributions listed
in (\ref{infs}) and (\ref{conts}).  That result describes the total
{\it local} anomaly on a particular $\a\b$-plane due to the
sources which we have described so far.  Since the total local anomaly
must vanish, there are two possibilities.  The first is
that the net anomaly vanishes identically.
The remaining possibility is that this net anomaly is non-zero, but is
cancelled by a contribution arising
from special couplings of local tensor fields which, thereby, provide
a local Green-Schwarz mechanism.  

Local tensor fields reside in twisted $N$=1 tensor multiplets 
on the $\a\b$-plane.  The two-form fields in such a multiplet
are anti-self-dual, in the sense that the associated 
gauge-invariant three-form field-strength satisfies $H=-*H$.
As a result of this, a special Chern-Simons interaction
$\int \d^{(5)}H Z_3^0$, where $\d^{(5)}$ is the five-form
brane-current with support on the $\a\b$-plane and 
$Z_4=dZ_3^0$ is a gauge-invariant four-form polynomial, will 
lead to two consequences.  First, due to the anti-self-duality, 
the electric coupling described by the Chern-Simons interaction
implies a magnetic coupling described by the Bianchi identity
$dH=Z_4$.  The magnetic coupling involves the same
polynomial as the electric coupling because anti-self-duality 
is equivalent to an electric-magnetic duality
\footnote{This is easy to
see by taking the exterior derivative of the Euler-Lagrange
equation $*H=-Z_3$, and then replacing $*H$ with $-H$.}.
Secondly, the Chern-Simons
coupling generates an inflow anomaly described by 
$I_8(GS)=Z_4\wedge Z_4$, where $GS$ designates Green-Schwarz.
Note that this polynomial is a perfect square
due to the duality. This is, in turn, 
a consequence of $N$=1 supersymmetry.  Thus, the net anomaly can be 
cancelled by specialized local tensor dynamics, provided that the net anomaly
reduces to a sum of perfect squares with one term for each available
tensor field.

\section{Local Anomaly Cancellation}

As a result of the above discussion, a program for analyzing the local
anomaly on a given $\a\b$-plane becomes apparent.  First, we 
assemble the net anomaly by summing the six terms in (\ref{infs})
and (\ref{conts}).  We will call this result $I_8$.
Then we sequence through the possibilities
$n_T$=0,1,2,..., in each case imposing that $I_8$ satisfies the
appropriate factorization requirement.  For the case $n_T$=0, 
since there are no local tensors, we require $I_8$=0.  For the 
case $n_T$=1, we impose that $I_8$ is proportional to a complete square, 
that is, $I_8\propto (Z_4)^2$.  
(In this second case, we simultaneously determine the form of the
electric and magnetic couplings of the local tensor.)   
For the case $n_T$=2, we require that $I_8$ is the sum of two perfect
squares.  And so forth.
These factorization requirements prove to be
marvelously restrictive.  For each choice of $n_T$, there results
unambiguous values for each and every one of the
previously unspecified geometric and topological parameters, including
the values of the magnetic charge, electric charges, and the identity 
of the gauge groups and representation content of the twisted sectors.

Since {\it M}-fivebranes
carry unit magnetic charge, as well as the zero mode 
fields described previously, we infer that
a fivebrane moving onto (or off of) a given $\a\b$-plane deposits
(or removes) charge and twisted modes in the process of doing so.
Since the fivebrane carries one $N$=1 tensor multiplet and one hypermultiplet,
we expect that the solutions to our factorization constraints will assemble
into hierarchies linked by incrementing the magnetic 
charge and the local tensor and hyper multiplicities as
$g\to g+1$, $n_T\to n_T+1$ and $n_H\to n_H+1$. This
is precisely what we find. Mathematically, this can be understood as follows.
A single fivebrane touching an orbifold fixed plane
is described by a singular object called a ``torsion free
sheaf'' \cite{si}. This sheaf carries one extra unit of magnetic charge and has
one tensor multiplet and one hypermultiplet as zero modes,  
identical to the analogous fivebrane data.
This accounts for the increase in each of these quantities by unity 
when the fivebrane is moved to a fixed plane.

At least in some cases, the torsion free sheaf can be shown to be the
singular ``small instanton'' limit of a smooth gauge instanton \cite{pt}. 
Smoothing out the sheaf into an instanton represents a true phase transition,
where the fivebrane data disappears and is replaced by a vector bundle
\cite{bopw}. 
In this process, the unit magnetic charge of the sheaf is
replaced by a unit increase in the second Chern number of the vector bundle.
The zero modes of the vector bundle are, in general, quite different from
those of the torsion free sheaf. Most importantly, the appearance of a smooth,
non-trivial vector bundle signals the breakdown
of the original twisted sector gauge group to a
smaller group ${\cal{G}}\subset E_8$. Therefore, after this phase
transition, we expect to have a smaller twisted sector gauge group with
identical topological charge but different numbers of tensor and
hypermultiplets.  Be that as it may, this theory remains anomaly free. 
As we will see, locally anomaly free orbifold planes do exist
that could be related to each other through small instanton phase
transitions. Thus, factorization of the local anomaly polynomial
yields an extra bonus. 

In addition to the torsion free sheaf 
transitions described earlier,
we expect another interesting grouping of our factorization
solutions.  In this second grouping, we expect small instanton transitions
between sets of solutions with identical local magnetic charge, but with different
numbers of zero modes and different local gauge groups.  

For clarity, let us recapitulate the fivebrane transitions discussed in the
previous two paragraphs. If a fivebrane moves to a particular 
$\a\b$-plane, it should
increment the local magnetic charge by one, and add one local
tensor and one local hypermultiplet to the associated twisted spectrum.
We would then attribute one unit of the total local magnetic charge to
the latent magnetic charge of the fivebrane, now interpreted
as a torsion-free sheaf.  The associated tensor
would be available to help mediate anomaly cancellation
through a local Green-Schwarz mechanism.  
Now, assume this configuration is a small instanton and can be deformed to a
smooth vector bundle. Then,
since the instanton does not have a tensorial zero mode,  
there would be one less tensor available to mediate the
anomaly cancellation.  The anomaly polynomial should, therefore, 
reconfigure so as to ensure continued anomaly cancellation,
but with a modified factorization criterion.
Thus, by classifying independent solutions to the factorization 
requirements into sets involving identical values of $g$ 
but different numbers of tensor fields,
one can {\it infer} such nontrivial phase transitions.  

We begin our analysis by considering a particular $\a\b$\ plane,
corresponding to one of the solid dots in Figure 1, representing
the unique six-dimensional intersection of a 
particular ten-dimensional $\a$-plane and a particular
seven-dimensional $\b$-plane.  To be more concrete, we
focus on one of the intersection points on the {\it lower} 
of the two $\a$-planes in
Figure 1, so that the local geometry is depicted as in Figure 2.
\begin{figure}
\begin{center}
\includegraphics[width=2.2in,angle=0]{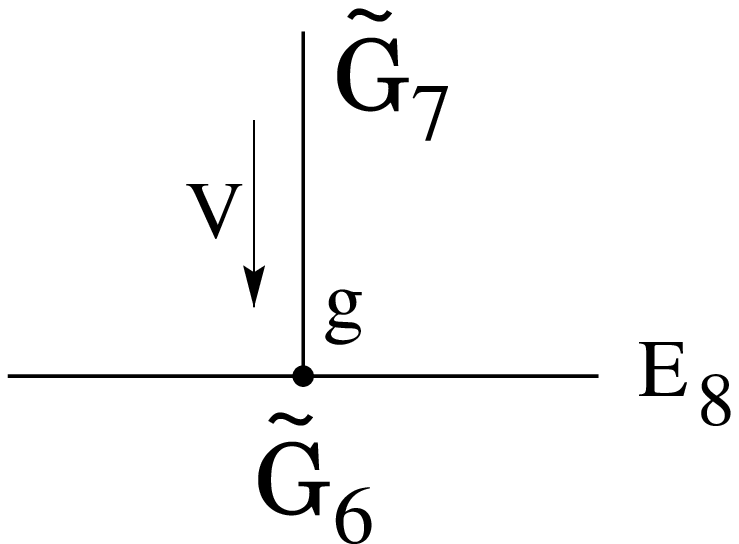}\\[.2in]
\parbox{4.5in}{Figure 2: The local geometry near a particular $\a\b$-plane,
showing some of the data to be resolved by anomaly factorization.}
\end{center}
\end{figure}
In Figure 2, the horizontal line represents the $\a$-plane, the vertical line 
represents the $\b$-plane and, finally,  the point of intersection 
represents the $\a\b$-plane.  The intersection
supports the local anomaly in which we are interested
and has magnetic charge $g$.  
The $\a$-plane supports $E_8$ super Yang-Mills fields, as described
above. This $E_8$ group is, in general, broken 
on the $\a\b$-plane to some subgroup depending of the action of $\b$
on the $E_8$ root lattice.  In the vertex-diagrams below, we indicate, 
to the right of the horizontal lines, 
the subgroup of $E_8$
left unbroken at the intersection.  Thus, Figure 2
indicates a scenario in which the full $E_8$ is left unbroken.  This figure
also indicates the presence of additional gauge structure with group
$\tilde{\cal{G}}_{7}$ localized
on the $\b$-plane and further
gauge structure on the $\a\b$-plane with group $\tilde{\cal G}_6$.  These
correspond to additional seven-dimensional and six-dimensional
fields, respectively.  

The only multiplet in seven dimensions is a vector multiplet,
which transforms in the adjoint of $\tilde{\cal G}_7$.
This decomposes into one six-dimensional $N$=1 vector multiplet and 
one six-dimensional hypermultiplet
\footnote{As a point of interest, this is the same 
decomposition enjoyed by a ten-dimensional vector multiplet.}.
An important fact is that the seven-dimensional fields are 
chirally projected by $\a$ onto the embedded six-dimensional 
$\a\b$-plane. It, thereby, contributes to the local $\a\b$ anomaly.

If $\tilde{\cal G}_7$ does not coincide with some factor of the broken
subgroup of $E_8$, then six-dimensional gauge invariance dictates
that the hypermultiplet is the part projected out by $\a$.  Thus,
the $\tilde{\cal G}_7$ gauge fields remain at the intersection
to enforce local $\tilde{\cal G}_7$ invariance.
The $\tilde{\cal G}_7$ adjoint gauginos fill out a
six-dimensional $N$=1 vector multiplet, which is indicated by the 
$V$ next to the downward arrow in Figure 2. This tells us that the 
``vector" part survives the $\a$ projection as we move down
the vertical line in that diagram and land on the intersection point.
Since the $N$=1 vector multiplet is chiral, the $\tilde{\cal G}_7$
gauginos will contribute to the $\a\b$ anomaly.
However, since this anomaly results from the coupling of fields
not localized on the $\a\b$-plane, this anomaly must be divided 
by two (since there are two $\a\b$-planes
embedded within a given $\b$-plane) compared to a similar anomaly due to 
six-dimensional $\tilde{\cal G}_7$ adjoint gauginos. 

If $\tilde{\cal G}_7$ coincides with a factor 
in the unbroken subgroup of $E_8$, then the $\tilde{\cal G}_7$ gauge fields 
on the $\a\b$-plane may be supplied by the ten-dimensional gauge
fields which survive the $\b$ projection.  
Consequently, the projection $\a$ should remove the
``vector" part of the seven-dimensional adjoint matter, so that 
the other part,  corresponding to an adjoint
hypermultiplet, survives the $\a$ projection.
We indicate this alternate situation by an $H$ next to the 
downward arrow in the corresponding diagram.  In this case, the
surviving {\it hyperinos} would contribute to the local anomaly.
This anomaly would include a division by two compared to a similar 
anomaly due to six-dimensional $\tilde{\cal G}_7$ adjoint hyperinos,
for reasons identical to those described in the preceeding paragraph
\footnote{The possibility of alternatively projecting out the 
vector or hypermultiplet parts of the seven-dimensional fields on
the six-dimensional planes was also mentioned in \cite{ksty}.}.

A third possibility is that the group $\tilde{\cal G}_7$ is
broken, by $\b$,
to some maximal subgroup ${\cal H}\subset \tilde{\cal G}_7$
on the $\a\b$-plane.  In this case, the seven-dimensional fields
would decompose into various six-dimensional fields transforming according
to representations determined by the appropriate branching rule.
These would include fields transforming in the adjoint of ${\cal H}$
and other fields transforming in other representations of ${\cal H}$.
The vector part of the adjoint fields would survive the $\b$ projection
while the hypermultiplet part of the remaining fields would survive.  
We indicate this hybrid situation by replacing the $V$ in Figure 2 with
the relevant subgroup 
${\cal H}\subset \tilde{\cal G}_7$ which survives the projection.  
 
In resolving the factorization criteria necessary to explain
local anomaly cancellation on a given $\a\b$-plane, one requires 
factors of two which can only be explained by
seven-dimensional matter in the manner we have just described.
Thus, the identity of seven-dimensional matter is indicated by
anomaly cancellation on an embedded sub-plane.  This is interesting 
because the seven-plane itself cannot support a local anomaly since
it is odd-dimensional.
(The situation is analogous to the fact that {\it eleven}-dimensional
supergravity is {\it needed} by the $E_8$ super-gauge multiplets on the
two $\a$-planes to render those ten-planes anomaly-free.)

There is one more subtle factor of one-half which needs
explanation.  This relates to hypermultiplets
in the {\it six-dimensional} twisted sector.  
It is possible for $r$ hypermultiplets to transform according to
a $2r$ dimensional representation $\tilde{\cal R}$
of the gauge group $\tilde{\cal G}$, provided the representation
is ``pseudoreal" in a sense to be clarified.  
In this case, the $4r$ scalar fields assemble into $r$ quaternions
represented as $\phi^\a_i$, where 
$i=1,2$ is an index which spans the {\bf 2} represention of an
$Sp(1)$ automorphism of the supersymmetry algebra
and $\a=1,...,2r$ spans $\tilde{\cal R}$.  
The group $\tilde{\cal G}$ acts as 
$\d\phi^\a_i=\t^a(T_a)^\a\,_\b$, where $(T_a)^\a\,_\b$
are antihermitian generators.  There 
exists a real invariant tensor $\rho^{\a\b}=\rho_{\a\b}$ 
which, by suitable field
redefinition,  can be put into  block-diagonal form 
$\rho={\rm diag}\,(i\s_2\cdots i\s_2\,)$, where
$\s_2$ is the second Pauli matrix.  The representation is 
pseudoreal if $(T_a^*)=-\rho\,T_a\,\rho$.  
See \cite{bsp,blp} for a more comprehensive discussion
of hypermultiplets.  In this case, we refer more
properly to $2r$ {\it half-hypermultiplets}, since the number of
hypermultiplets is half the dimensionality of the representation.
Each such half-hypermultiplet then contributes one-half of the anomaly 
which we would normally attribute to $2r$ antichiral spinors 
transforming in $\tilde{\cal R}$ via naive application of index theorems.   

Next, we should explain how to algebraically characterize the possible
branching patterns describing the projection of the two $E_8$ factors by 
$\b$.  The simplest possibility is the one indicated in Figure 2, 
where the relevant $E_8$ factor remains unbroken.  In terms of 
six-dimensional $N$=1 multiplets, the ten-dimensional $E_8$ 
vector multiplet decomposes into one vector multiplet and one hypermultiplet.
In this case, it is the hypermultiplet components which are
projected out by $\b$ on the $\a\b$-plane.  This leaves us with the
gauge fields necessary to enforce local $E_8$ invariance on the 
$\a\b$-plane.  This also fixes two of our parameters, $(V_B,H_B)=(248,0)$.  
In this case, all $E_8$ traces which appear in both 
the inflow anomalies (\ref{infs}) and the quantum anomalies 
(\ref{conts}) are each taken at face value.  Thus, using the representation
indices in Table 1, we can use the results
${\rm Trace}_{\bf 248}\,F^2\equiv 30\,{\rm tr}\,F^2$
and ${\rm Trace}_{\bf 248}\,F^4=9({\rm tr}\,F^2)^2$ to express the 
traces which appear in the quantum anomalies (\ref{conts}) in terms 
of the fundamental (${\rm tr}$) traces
\footnote{Note that the traces which appear in
the inflow anomaly (\ref{infs}), through the implicit dependence 
of (\ref{polys}), are fundamental traces to begin with.}.  
 
More generally, the $E_8$ factors will be broken by $\b$, on
the $\a\b$-planes, to some maximal subgroup with a branching 
pattern which can be found from the tables in
\cite{slansky}.   In this case, when we determine our anomaly polynomial 
$I_8$ by adding up the six contributions in (\ref{infs}) and (\ref{conts}),
we replace the various $E_8$ traces by traces over the relevant 
representations of the residual subgroup.  We then relate these to
traces over fundamental representations of the factors in this subgroup
by using representation indices, such as those listed in Table 1.
However, there is a subtle difference between the inflow contributions
(\ref{infs}) and the quantum contributions (\ref{conts})
which should be properly accounted for, and which we now describe.  
Since the inflow anomalies are classical expressions, we can apply the 
group theoretic reduction directly on the traces which appear in 
(\ref{infs}).  However, in the general case, 
the six-dimensional quantum anomalies derive from both
chiral and antichiral fields. The chiral fields, which satisfy 
$\Gamma_7\psi=\psi$, appear in the six-dimensional $N$=1 vector
multiplets.  The antichiral fields, which satisfy $\Gamma_7\psi=-\psi$,
occur in hypermultiplets.  Since chiral and antichiral
fields contribute one-loop anomalies with opposite sign, there is an extra
minus sign associated with all quantum anomalies arising from hypermultiplet
couplings.  We illustrate this with two explicit examples.
 
As a first example, we choose the breaking pattern $E_8\to E_7\times SU(2)$.
In this case, we have the branching rule
${\bf 248}=({\bf 133},{\bf 1})\oplus({\bf 1},{\bf 3})\oplus 
({\bf 56},{\bf 2})$. We determine that the surviving six-dimensional
fields comprise 133 $N$=1 vector multiplets transforming as the adjoint
of $E_7$, another three vector multiplets 
transforming as the adjoint of $SU(2)$ and
112 hypermultiplets transforming as a bifundamental representation.  
Thus, $(V_B,H_B)=(136,112)$.
In this case, we reduce the $E_8$ traces which occur
in the inflow anomaly as follows
\brr {\rm tr}F^2 &=& \ft{1}{30}\,{\rm Tr}_{\bf 248}\,F^2
     \nonumber\\[.1in]
     &=& \ft{1}{30}\,\bpl\,
     {\rm Tr}_{\bf 133}\,F_a^2+2\,{\rm tr}_{\bf 56}\,F_a^2
     +{\rm Tr}_{\bf 3}\,F_b^2+56\,{\rm tr}_{\bf 2}\,F_b^2\,\bpr
     \nonumber\\[.1in]
     &=& \ft16\,{\rm tr}\,F_a^2+2\,{\rm tr}\,F_b^2 \,,
\label{e7su2}\err
where we have used the indices in Table 1.
In (\ref{e7su2}), the subscripts $a$ and $b$ denote 
$E_7$ and $SU(2)$ respectively.
In the final line, we have dropped the labels from the fundamental 
${\bf 56}$ and ${\bf 2}$ traces.  
Thus, to describe the inflow anomaly in the case where 
the $E_8$ factor is broken by $\b$ to $E_7\times SU(2)$, 
we substitute the identity (\ref{e7su2}) for the ${\rm tr}\,F^2$ 
in the inflow anomaly (\ref{infs}).
In the quantum anomaly, on the other hand, 
hypermultiplets and vector multiplets contribute with opposite signs.  
As a result, when computing the local one-loop anomaly 
in the case where $E_8\to E_7\times SU(2)$, we
we should replace the factor ${\rm trace}\,F^2$ which appears in
$I^{(1/2)}_{{\rm MIXED}}$ using the
following
\footnote{See equation (C.2) of \cite{mlo} for the explicit 
polynomial corresponding to $I_{MIXED}^{(1/2)}$, as well as all
of the other quantum anomaly polynomials referred to in this paper.}
\brr {\rm trace}F^2 &=& 
      {\rm Tr}_{\bf 248}\,F^2
     \nonumber\\[.1in]
     &=&
     {\rm Tr}_{\bf 133}\,F_a^2-2\,{\rm tr}_{\bf 56}\,F_a^2
     +{\rm Tr}_{\bf 3}\,F_b^2-56\,{\rm tr}_{\bf 2}\,F_b^2
     \nonumber\\[.1in]
     &=& {\rm tr}\,F_a^2-52\,{\rm tr}\,F_b^2 \,.
\label{qq}\err
This derivation differs from (\ref{e7su2}) by the minus signs 
on terms relating to hypermultiplet couplings.   
As described above, these minus signs reflect the 
antichirality of hyperinos. 
Similar comments apply to the term ${\rm trace}\,F^4$ which 
appears in $I^{(1/2)}_{\rm GAUGE}$.  In total, 
to describe the case $E_8\to E_7\times SU(2)$ 
we should make the following replacements
\brr {\rm tr}F^2 &=&  
     \ft16\,{\rm tr}\,F_a^2+2\,{\rm tr}\,F_b^2 
     \nonumber\\[.1in]
     {\rm trace}\,F^2 &=&  
     {\rm tr}\,F_a^2-52\,{\rm tr}\,F_b^2 
     \nonumber\\[.1in]
     {\rm trace}\,F^4 &=&  
     \ft{1}{12}\,({\rm tr}\,F_a)^2
     -20\,({\rm tr}\,F_b)^2
     -6\,{\rm tr}\,F_a^2\wedge {\rm tr}\,F_b^2 \,.
\label{a}\err
The first of these should be substituted in the classical (inflow)
anomaly, while the second two should be substituted in the quantum anomaly.
Note our mnemonic that traces which appear in the quantum anomaly
are designated ``trace", whereas classical traces
are abbreviated ``tr".

As a second example, we choose the breaking pattern
$E_8\to SO(16)$.  In this case, we have the branching rule
${\bf 248}\to {\bf 120}\oplus {\bf 128}$.
We determine that the surviving six-dimensional
fields comprise 120 $N$=1 vector multiplets transforming as the adjoint
and 128 hypermultiplets transforming as the spinor of $SO(16)$.
Thus, $(V_B,H_B)=(120,128)$. 
In this case, we reduce the $E_8$ traces which occur
in the inflow anomaly and in the quantum anomaly
in a manner similar to that described in our previous example,
making use of the representation indices in Table 1. The appropriate
reductions are
\brr {\rm tr}F^2 &=&  
     {\rm tr}\,F_a^2 
     \nonumber\\[.1in]
     {\rm trace}\,F^2 &=&  
     -2\,{\rm tr}\,F_a^2 
     \nonumber\\[.1in]
     {\rm trace}\,F^4 &=&  
     -3\,({\rm tr}\,F_a)^2
     +16\,{\rm tr}\,F_a^4 \,,
\label{b}\err
where the subscript $a$ now denotes $SO(16)$.  
Once again, the first of these should be substituted in the classical (inflow)
anomaly, while the second two should be substituted in the quantum anomaly.

Using the three distinct cases which we have so far addressed,
corresponding to the choices where $\b$ breaks
$E_8$ to $E_8$, $E_7\times SU(2)$ or $SO(16)$, we have enough data 
to completely determine an interesting set of solutions to our anomaly
factorization problem. These solutions conform to our expectations by 
assembling into hierachies as described previously.

On a given $\a\b$-plane, such as that depicted by the
intersection point in Figure 1, 
the net six-dimensional anomaly $I_8$
is determined by adding up all six terms in (\ref{infs})
and (\ref{conts}). One then substitutes  identities, such as (\ref{a}) 
and (\ref{b}), relevant to the particular $E_8$ breaking pattern
being considered, in the manner explained above.
What results is a polynomial with terms proportional to the each of 
${\rm tr}\,R^4$,
$({\rm tr}\,R^2)^2$,
${\rm tr}\,R^2\wedge {\rm tr}\,F^2$,
${\rm tr}\,R^2\wedge {\rm tr}\,F^2$,
$({\rm tr}\,F^2)^2$,
${\rm tr}\,F^2\wedge {\rm tr}\,{\cal F}^2$,
${\rm tr}\,F^4$, and
${\rm tr}\,{\cal F}^4$, where
$F$ stands generically for factors in the residual
group ${\cal G}\subset E_8$ which survives the $\b$ projection
(we have denoted these $F_a$ and $F_b$ above) and ${\cal F}$
stands generically for factors in any gauge group associated with
twisted matter.

Note that we have allowed for twisted fields which are either 
six or seven dimensional.  In the former case, we refer to $N$=1 
fields living exclusively on the $\a\b$-plane under consideration.
In the latter, we refer to vector adjoint supermultiplets living on
the seven-dimensional $\b$-plane which intersects this $\a\b$-plane.
The seven-dimensional fields will contribute to the anomaly with a tell-tale
factor of two, as described above.  
Computationally, we accomodate both of these cases
simultaneously by formally allowing $n_T$ tensors, $n_V$ vectors and $n_H$
hypermultiplets, where $n_V$ and $n_H$ can assume half-integral
values.   The (formal) appearance of half-integer numbers of
multiplets then indicates that the associated matter is seven-dimensional.
 
Keeping the twisted matter arbitrary, we determine $I_8$ by
adding up all six contributions in (\ref{infs}) and (\ref{conts}).
For any choice of ${\cal G}\subset E_8$, we reduce the
various $E_8$ traces to ${\cal G}$ traces according to the scheme 
described above.  
This process provides us with a provisional form of the local
anomaly.  It remains provisional since the twisted contribution
remains to be resolved.  Nevertheless, we can extract our first bits
of useful information.  Since anomaly cancellation is possible
only if $I_8$ either vanishes identically or reduces to a sum of
perfect squares, it follows that any nonfactorizable terms 
in $I_8$ must vanish.  The vanishing of the ${\rm tr}\,R^4$ term requires
\brr n_H-n_V=30g-29n_T+\ft{1}{2f}\,(244-h)+\ft{1}{f}(V_B-H_B) \,.
\label{nnn}\err
This constraint is the local version of the global constraint
$N_H-N_V+29 N_T=273$, where $N_H$, $N_V$ and $N_T$ are the 
{\it total} number of hyper, vector and tensor multiplets in the 
entire orbifold, including all twisted and untwisted contributions.  
The relationship between the local constraint (\ref{nnn}) and the
global version is described in \cite{mlo}.  Note that (\ref{nnn})
is invariant when $g\to g+1$, $n_T\to n_T+1$, and $n_H\to n_H+1$,
consistent with expectations described previously.
 
Henceforth, we concentrate on the 
$S^1/{\bf Z}_2\times T^4/{\bf Z}_2$ orbifold.  In this case, 
we have $(f,h)=(16,4)$, as described above, so that 
(\ref{nnn}) becomes
\brr n_H-n_V=30g-29n_T+\ft{15}{2}+\ft{1}{16}\,(V_B-H_B) \,.
\label{ee}\err
Since the left-hand side must be either an integer
or a half-integer, it follows that $(V_B-H_B)/16$
must be integer or half-integer as well, 
since in this case $g$ is quantized in quarter-integer units.
Each of the three $E_8$ breaking patterns which we have addressed,
${\cal G}=E_8, E_7\times SU(2)$ and $SO(16)$,
corresponding to $(V_B,H_B)=(248,0), (136,112)$ and $(120,128)$
respectively, respect this constraint.  We restrict our
study to these three breaking patterns.

The sytematic analysis of local anomaly cancellation proceeds
as outlined above. We first seek solutions with no twisted
tensor multiplets, so that $n_T=0$.  In this case, $I_8$ 
must vanish identically.
We consider each of the three possible $E_8$
breaking patterns described above, using the relevant values
of $V_B$ and $H_B$ in each case.  For these three 
possibilities, when $n_T=0$ equation (\ref{ee}) reduces to the
constraints indicated in Table 2.
\begin{figure}
\begin{center}
\begin{tabular}{c|c}
\hspace{1in} & \hspace{1in} \\[-.2in]
${\cal G}$ & $n_H-n_V$ \\[.1in]
\hline
$E_8$ & $30g+23$ \\[.1in]
$E_7\times SU(2)$ & $30g+9$ \\[.1in]
$SO(16)$ & $30g+7$ \\[.1in]
\end{tabular}\\[.15in]
\parbox{4in}{Table 2: Constraints linking the local magnetic
charge and the multiplicities of twisted hyper and vector multiplets
when $n_T$=0 for three possible $E_8$ breaking patterns.} 
\end{center}
\end{figure}
Thus, if there are no local twisted tensor multiplets, 
anomaly cancellation implies the indicated correlations
between the local magnetic charge and the multiplicities
of twisted hyper and vector multiplets.  Note that, in each case,
extra twisted matter is required since the indicated
multiplicites can not be made to vanish with any properly 
quantized choice of $g$.
The challenge in the case $n_T$=0 is not only to identify the
multiplicities of twisted states, but also to 
identify the twisted gauge groups, the representation content
of the twisted matter,   
as well as values of $g$, $\eta$ and $\rho$
which satisfy the restrictions in
Table 2 (which ensures cancellation of the local ${\rm tr}\,R^4$ 
anomaly). Furthermore, these choices must provide for complete cancellation 
of all other terms in the full polynomial $I_8$.  Satisfying all of these 
requirements is a highly restrictive demand.

We first look for a ``basic" solution where ${\cal G}=E_8$. 
In this case, we find a unique solution to all of our constraints.
This solution requires $g=-3/4$ and 
$\tilde{\cal G}_7=SU(2)$, which is broken as $SU(2)\to U(1)$ on the
$\a\b$-plane.  In this case, the three adjoint $SU(2)$ fields
provide one six-dimensional vector multiplet and two 
hypermultiplets on the $\a\b$ plane.  There are no further twisted
fields. Thus, the gauge structure on the $\a\b$-plane is $E_8\times U(1)$,
under which the twisted fields transform as follows
\vspace{.1in}
\begin{center}
\begin{tabular}{ccl}
Vectors: & $\ft12\,{\bf 1_{(0)}}$ & $n_V=1/2$ \\[.1in]
Hypers: &  $\ft12\,{\bf 1_{(+1)}}\oplus 
\ft12\,{\bf 1_{(-1)}}$ & $n_H=1/2+1/2=1$ \\[.1in]
Tensors: & None & $n_T=0$ \,,
\end{tabular}
\end{center}
\vspace{.1in}
where the $U(1)$ charges are indicated in the parenthetical subscripts.
This solution also requires $(\eta,\rho)=(1,0)$
\footnote{The requirement that $\rho=0$ and the need for $U(1)$ 
gauge factors in the ``basic" solution to 
$S^1/{\bf Z}_2\times T^4/{\bf Z}_2$ orbifolds
was also discussed in \cite{ksty}.}.
In this case, the anomaly vanishes completely, as it should
since there are no tensors to provide a local Green-Schwarz
mechanism.  The factor of one-half accompanying the twisted
field representions indicate that these describe 
{\it seven}-dimensional fields
living on the $\b$-plane, contributing via a chiral
projection onto the embedded $\a\b$-plane.  
We represent this local solution 
with the diagram shown in Figure 3.
\begin{figure}
\begin{center}
\includegraphics[width=2.5in,angle=0]{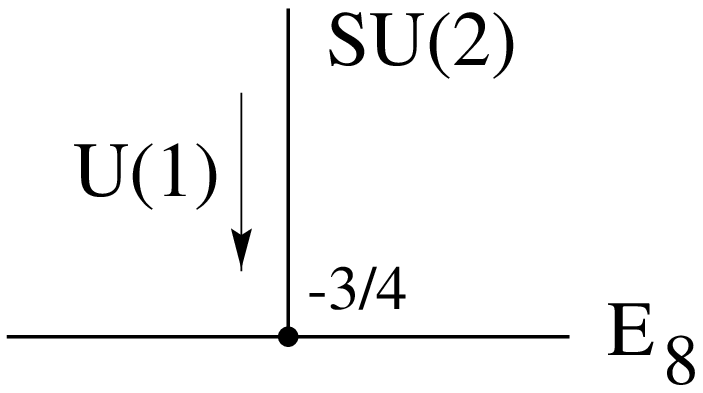}\\[.2in]
\parbox{4.3in}{Figure 3: Vertex diagram corresponding to
the ``basic" solution in the $S^1/{\bf Z}_2\times T^4/{\bf Z}_2$
orbifold, as described in the text.}
\end{center}
\end{figure}
In Figure 3, the twisted gauge group $SU(2)$ is indicated next to the
vertical line, signifying that this corresponds to $\tilde{\cal G}_7$.
The magnetic charge associated with
the intersection is indicated by the $-3/4$. There is no local $U(1)$
anomaly because the charges cancel, leaving a net 
$U(1)$ charge of zero.
 
Our basic solution has a two notable aspects.
First, using the multiplicities listed above, we compute 
$n_H-n_V=1/2$, which is precisely the value specified in Table 2
for the case of unbroken $E_8$ and for the choice $g=-3/4$.   
The second notable aspect 
concerns the magnetic charge.  For reasons described in \cite{mlo},
we attribute a topological significance to this number. Specifically,
$g$ corresponds to an $E_8$ instanton number (associated with
an instanton residing on the $\a\b$-plane) minus 
the local contribution due to the nontrivial Euler character of 
the $K3$ manifold. (We attribute the second of these to
a local gravitational instanton, associated with a pointlike version 
of the ALE space needed to blow up the orbifold.)  Since the Euler number
of $K3$ is 24, we divide this evenly over the 32 $\a\b$-planes, so 
that the local gravitational contribution to $g$ should be exactly
$-3/4$. Since we are considering the case of unbroken $E_8$, we assume there
are no local gauge instantons. Therefore, the only
contribution to $g$ should be the gravitational result
of $-3/4$. It is gratifying that this number is required by our independent 
anomaly cancellation requirements.

When we impose $n_T$=0 and $I_8=0$ on the cases 
where $\b$ breaks $E_8$ to ${\cal G}=E_7\times SU(2)$ 
or ${\cal G}=SO(16)$, we also find unique solutions with specialized
values of $g$ and with specific twisted matter content.
These solutions are described by the diagrams in the left-hand column 
of Figure 4.  Note that, in 
each case, we require seven-dimensional $SU(2)$ vector multiplets.  
In the ${\cal G}=E_7\times SU(2)$ case, 
the seven-dimensional $SU(2)$ factor is identified with
the $SU(2)$ factor in ${\cal G}$.  This identification is indicated
by the asterix on the two $SU(2)$ factors in the relevant diagram. 
As described previously, under
these circumstances, $\a$ projects out the ``vector" component
of the seven-dimensional matter but preserves the ``hyper"component.
This is represented by the $H$ next to the arrow in the same diagram.
In contrast to our ``basic" solution, both of these new solutions 
involving local breakdown of $E_8$ 
require $(\eta,\rho)=(1,1)$.  The nonvanishing of $\rho$
indicates that the $\b$-planes 
support non-zero $SU(2)$ electric charge in these cases. 
The physics of this observation may prove interesting.
\begin{figure}
\begin{center}
\begin{tabular}{ll}
\hspace{2.7in} & \hspace{2.7in} \\[1in]
\includegraphics[width=2.3in,angle=0]{e80.eps} &
\includegraphics[width=2.3in,angle=0]{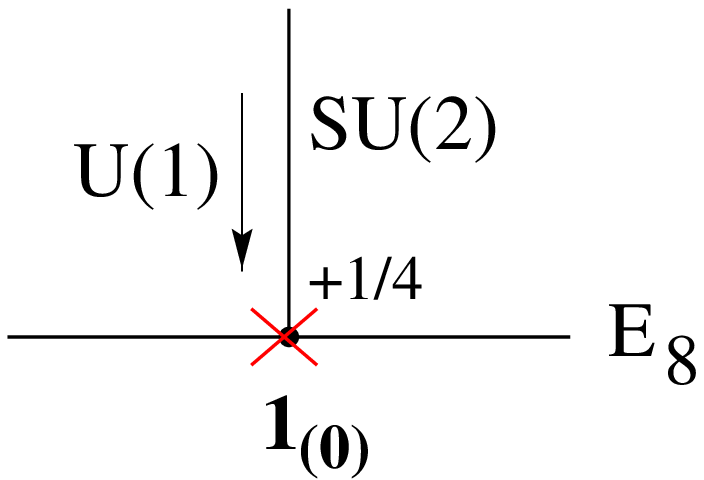} \\[1in]
\includegraphics[width=2.7in,angle=0]{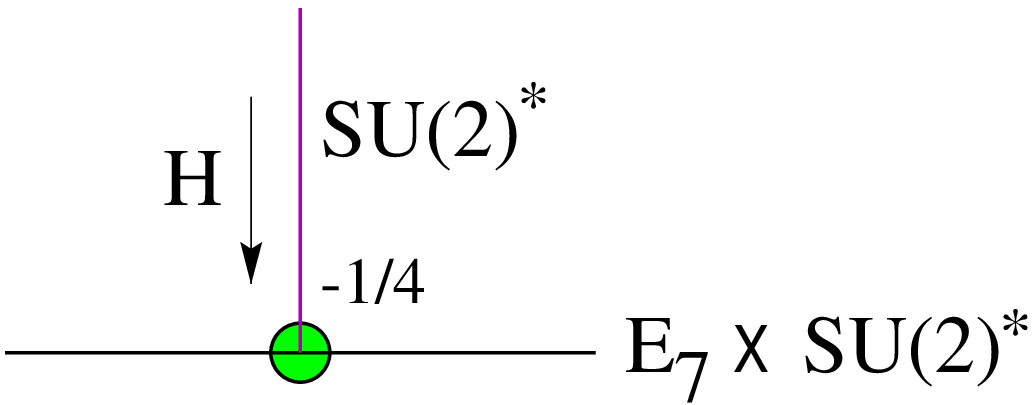} &
\includegraphics[width=2.7in,angle=0]{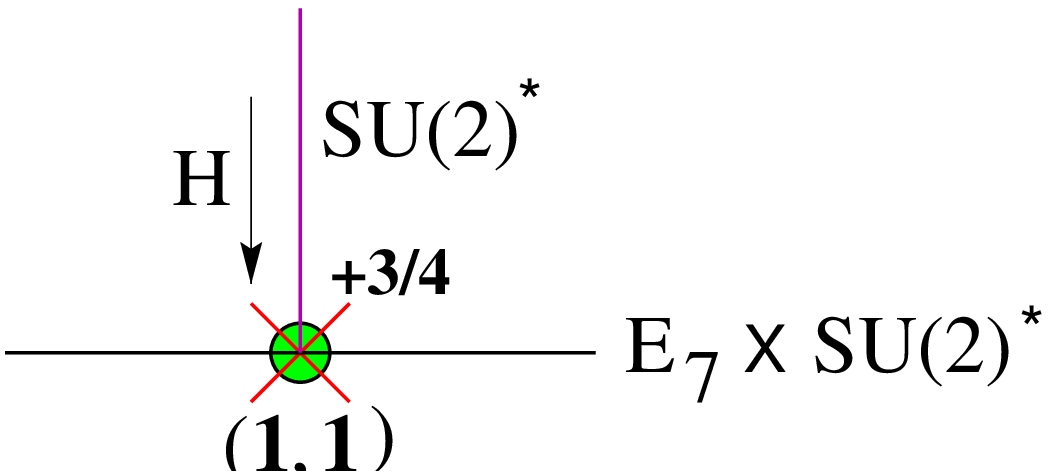} \\[1in]
\includegraphics[width=2.6in,angle=0]{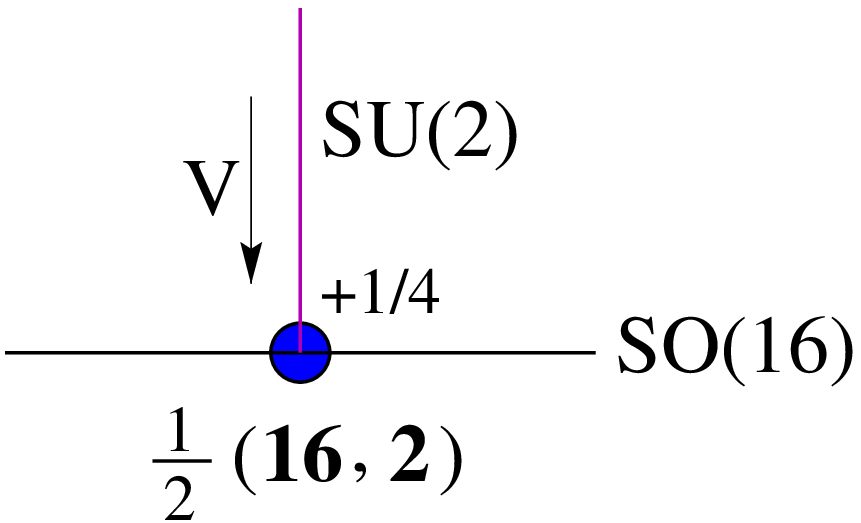} &
\includegraphics[width=2.6in,angle=0]{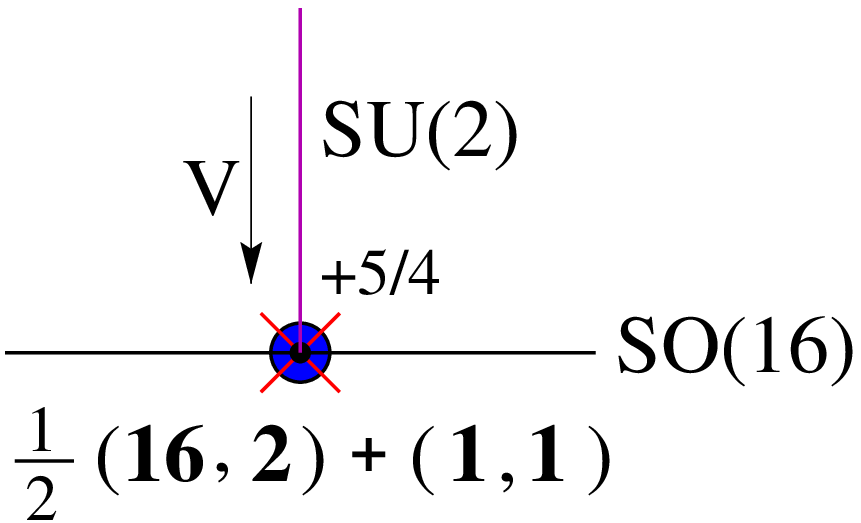}
\end{tabular}\\[.5in]
\parbox{6.3in}{Figure 4:  A collection of consistent orbifold vertices
for the $S^1/{\bf Z}_2 \times T^4/{\bf Z}_2$ orbifold for three possible
$E_8$ breaking patterns and for the two possibilities $n_T$=0 and
$n_T$=1.  In the middle two diagrams, the asterix signifies that the
indicated $SU(2)$ factors are identified.}
\end{center}
\end{figure}

We interpret the $\b$-induced $E_8$ symmetry breakdown as a reflection of 
instantons residing on the $\a\b$-plane.  Since there
are two classes of gauge instantons which could reside on the
$\a\b$-planes, one contributing integer magnetic charge
and the other contributing half-integer magnetic charge, we infer
that solutions could exist with $g$ taking values at half-integer
increments greater than the ``basic" value of $-3/4$.  This is
precisely what we find; the $n_T$=0 solutions for 
${\cal G}=E_7\times SU(2)$ and ${\cal G}=SO(16)$ require
$g=-1/4$ and $g=+1/4$, corresponding to half-integer
and integer valued instantons, respectively.
\begin{figure}
\begin{center}
\begin{tabular}{c|c}
\hspace{1in} & \hspace{1in} \\[-.2in]
${\cal G}$ & $n_H-n_V$ \\[.1in]
\hline
$E_8$ & $30g-6$ \\[.1in]
$E_7\times SU(2)$ & $30g-20$ \\[.1in]
$SO(16)$ & $30g-22$ \\[.1in]
\end{tabular}\\[.15in]
\parbox{4in}{Table 3: Constraints linking the local magnetic
charge and the multiplicities of twisted hyper and vector multiplets
when $n_T$=1 for three possible $E_8$ breaking patterns.} 
\end{center}
\end{figure}
 
In the next phase of our systematic search for local anomaly-free
vertices, we study the cases with $n_T$=1.  
In these cases, we impose that the anomaly factorizes
as a complete square, $I_8\propto (Z_4)^2$, so that it
can be cancelled by dynamics involving the self-dual tensor
in the local twisted spectrum. 
Equation (\ref{ee}), which enforces the
vanishing of the ${\rm tr}\,R^4$ term in $I_8$,
is still important since ${\rm tr}\,R^4$
cannot factorize.  Once again, we consider each of the three
$E_8$ breaking patterns discussed above, using the relevant values of
$V_B$ and $H_B$ in each case.  For these three possibilities,
when $n_T$=1, equation (\ref{ee}) reduces to the constraints indicated
in Table 3.  
Thus, if there is one local twisted tensor multiplets then
local anomaly factorization implies the indicated correlations
between the local magnetic charge and the multiplicities
of twisted hyper and vector multiplets.  Note that, in each case,
extra twisted matter is still required since the indicated
multiplicites can not be made to vanish with any properly 
quantized choice of $g$.
The challenge, in the case $n_T$=1, is not only to identify
multiplicities of twisted states, but also 
the twisted gauge groups, the represention content of the twisted fields,  
as well as values of $g$, $\eta$ and $\rho$
which can satisfy the restrictions in
Table 3 (which ensures cancellation of the local ${\rm tr}\,R^4$ 
anomaly). Furthermore, they must provide the appropriate factorization of $I_8$.  
Satisfying these requirements is, again, a highly restrictive demand.

As before, we start
with the case where $E_8$ remains unbroken.
We again find a unique solution to our
constraints, but this time with $g=+1/4$.  Once again 
$\tilde{\cal G}_7=SU(2)$, and the seven dimensional
gauge group is broken by $\b$ on the $\a\b$-plane as
$SU(2)\to U(1)$.  The only change to the twisted spectrum in the
analogous $n_T=0$ solution is the addition of one singlet
hypermultiplet to the local {\it six}-dimensional spectrum.
Thus, the twisted fields transform under $E_8\times U(1)$ as follows 
\vspace{.1in}
\begin{center}
\begin{tabular}{ccl}
Vectors: & $\ft12\,{\bf 1_{(0)}}$ & $n_V=1/2$ \\[.1in]
Hypers: &  $\ft12\,{\bf 1_{(+1)}}\oplus 
\ft12\,{\bf 1_{(-1)}}\oplus {\bf 1}_{(0)}$ & $n_H=1/2+1/2+1=2$ \\[.1in]
Tensors: & ${\bf 1}_{(0)}$ & $n_T=0$ \,,
\end{tabular}
\end{center}
\vspace{.1in}
where the $U(1)$ charges are again indicated in the parenthetical subscripts.
This solution also requires $(\eta,\rho)=(1,0)$
The factors of one-half indicate that these fields describe a
projection of a seven-dimensional
multiplet, living on the $\b$-fixed plane, via a chiral
projection onto its boundary.  This solution is shown in the 
upper right-hand diagram in Figure 4.  In our diagrams, 
we indicate the presence of a twisted tensor multiplet 
by an \includegraphics[width=.15in,angle=0]{x.eps}.
In this case, the anomaly does not vanish, but is given by
\brr I_8=-\frac{1}{(2\pi)^34\!}\,\frac{3}{16}\,
     \bpl\,{\rm tr}\,R^2-2\,{\rm tr}\,F^2\,\bpr^2 \,.
\err
Since this is proportional to a perfect square, 
it can be removed by a local Green-Schwarz
mechanism mediated by the anti-self-dual tensor in the twisted tensor 
multiplet.

When we impose $n_T$=1 and $I_8\propto(Z_4)^2$ on the cases 
where $\b$ breaks $E_8$ to ${\cal G}=E_7\times SU(2)$ 
or ${\cal G}=SO(16)$, we also find unique solutions with specialized
values of $g$ and specific twisted matter content.
The set of $n_T$=1 solutions corresponding to each of the
our three choices for ${\cal G}$ are described by the 
diagrams in the right-hand column of Figure 4. 
In each of these cases, the presence of a twisted tensor multiplet is 
indicated by the \redx\, on the relevant vertex.
In these cases,
the anomaly $I_8$ does not vanish but, rather, is given by a complete square.
Therefore, the twisted tensor involves interesting dynamics.
Notably, the cases where $E_8$ is broken require 
$\rho=1$.  This is in contrast to the situation involving unbroken $E_8$, 
where $\rho=0$.

It is useful to compare the top right
vertex-diagram in Figure 4 with the lower left vertex-diagram 
in that same figure.  Since these
two vertices have identical magnetic charge, we infer 
a transition whereby the fivebrane connects smoothly to an instanton,
locally breaking $E_8$ to $SO(16)$.  A subtle point concerns the
electric charge of the associated seven-dimensional $\b$-plane
(the vertical line in these diagrams). 
In the unbroken ($E_8$) phase we require $\rho=0$, so the seven-plane 
is not electrically charged. However, in the broken 
($E_7\times SU(2)$ or $SO(16)$) phases
we require $\rho=1$.  This process is shown by the two-diagram
sequence depicted in Figure 5.
\begin{figure}
\begin{center}
\begin{tabular}{ll}
\hspace{2.7in} & \hspace{2.7in} \\[1in]
\includegraphics[width=2.3in,angle=0]{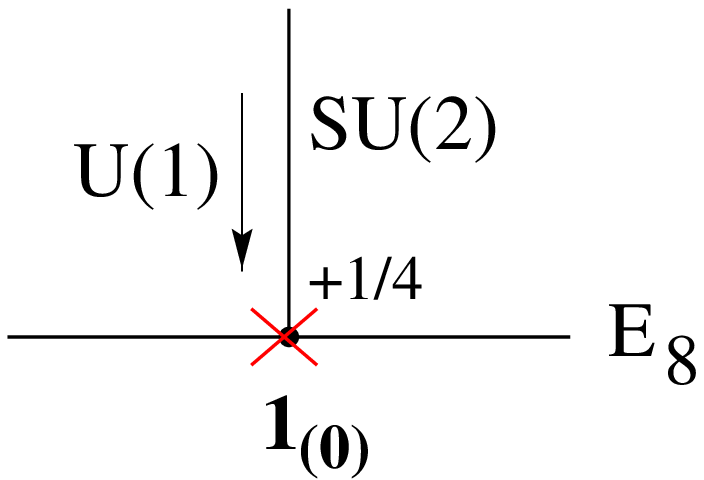} &
\includegraphics[width=2.6in,angle=0]{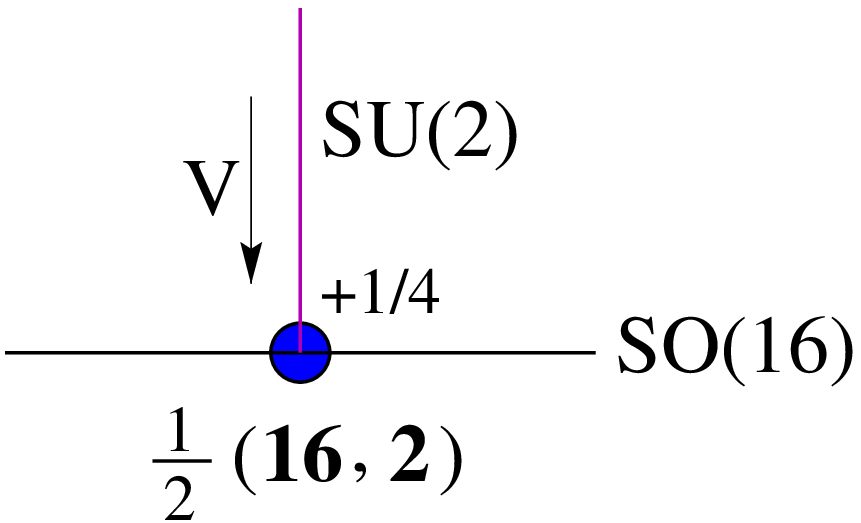} 
\end{tabular}\\[.2in]
\parbox{5in}{Figure 5: A two step process in which a fivebrane 
which has moved to a vertex smoothly deforms to an integrally-charged
instanton.}
\end{center}
\end{figure}

\section{The Global Structure}

Now that we have tabulated some consistent local solutions, as
classified in Figure 4, we can attempt to assemble these 
into a coherent global orbifold.  There are extra
constraints on this procedure, however, which need to be taken into
account.  The first of these 
is implied by the exactness of $dG$ and follows from
integrating the Bianchi 
identity (\ref{bianchi}) over the five compact dimensions.
Since this region has no boundary, the left-hand side
of the integrated version of (\ref{bianchi}) vanishes due to
Stokes theorem since the integrand is a total derivative.
This implies that the net magnetic charge of the
entire orbifold is zero.  Without loss of generality, we can
concentrate all of the magnetic sources either on the $\a\b$-planes
or on fivebrane worldvolumes.   
We therefore determine that
\brr N_5+\sum_{i=1}^{32}g_i=0 \,,
\err
where $N_5$ is the number of fivebranes not residing on
$\a\b$-planes.  Note that $N_5$ is necessarily a positive integer.

There is a unique ``basic" global configuration 
satisfying this contraint for which neither of the $E_8$ factors is
broken at any intersection.  In this case, each $\a\b$-plane carries a
magnetic charge of $-3/4$. Since there are thirty-two $\a\b$-planes,
the proper magnetic balance
is minimally achieved by including 24 fivebranes distributed randomly
in the bulk of the orbifold.  This situation is depicted by the first diagram
in Figure 6.
\begin{figure}
\begin{center}
\includegraphics[width=2.6in,angle=0]{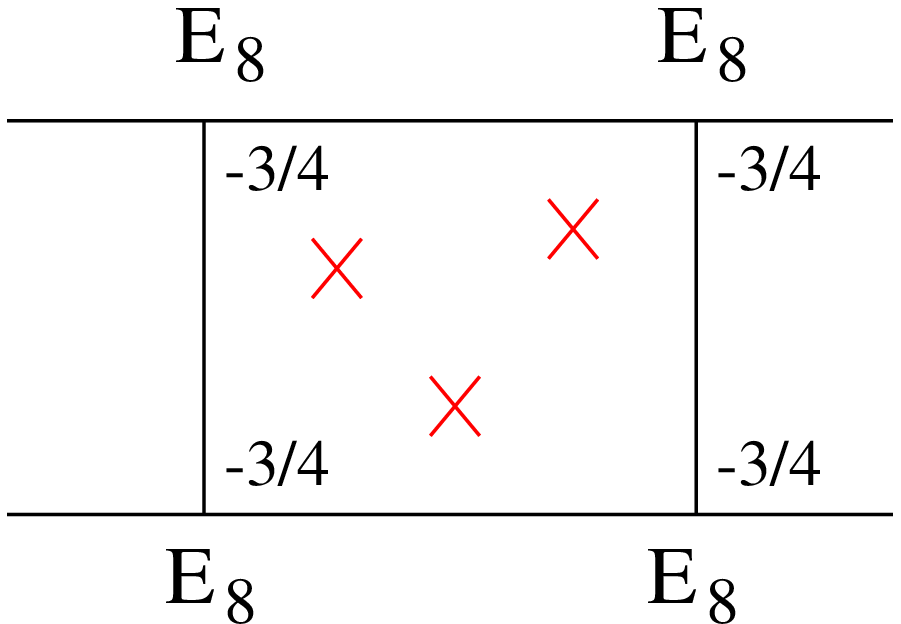}\\[.2in]
\includegraphics[width=2.6in,angle=0]{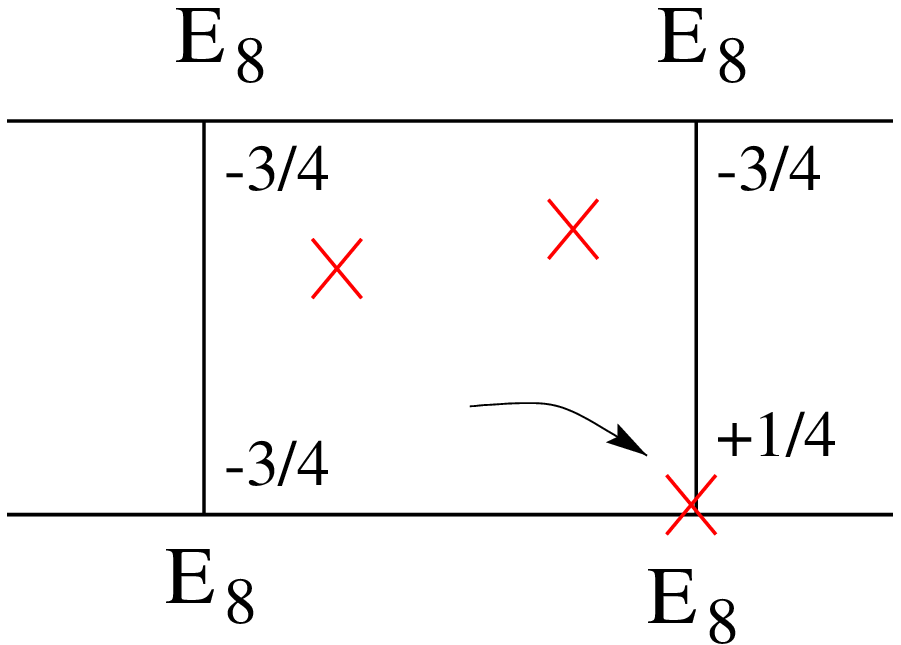}\\[.2in]
\includegraphics[width=2.6in,angle=0]{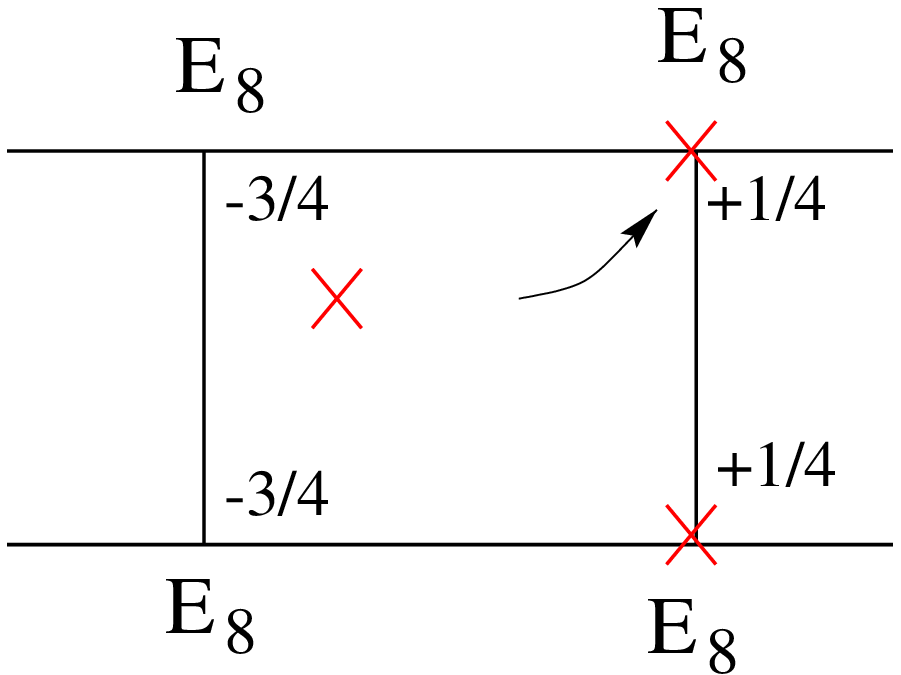}\\[.2in]
\includegraphics[width=2.6in,angle=0]{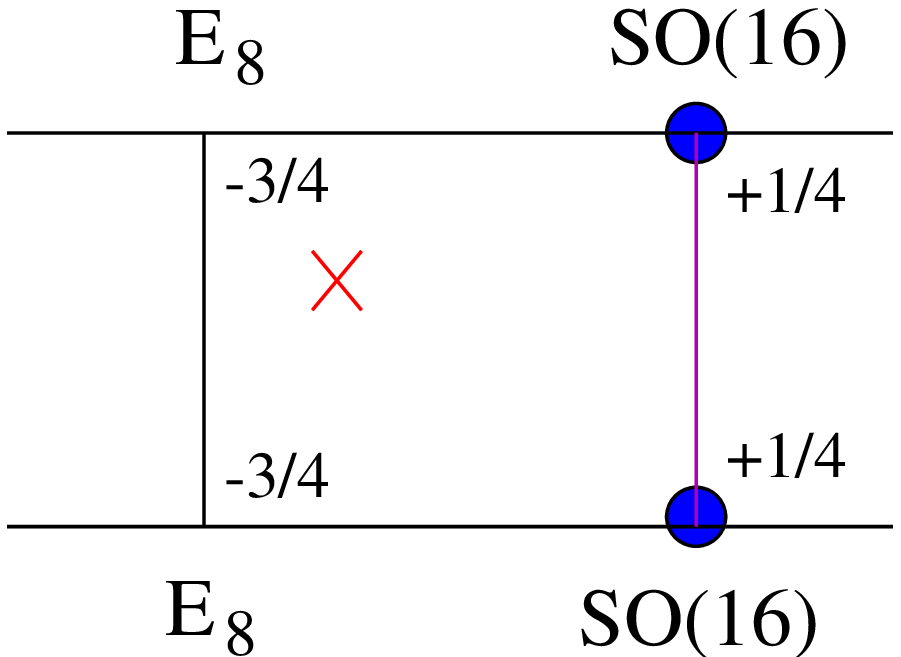}\\[.2in]
\parbox{4in}{Figure 6: A global picture of a phase transition in the
$S^1/{\bf Z}_2\times T^4/{\bf Z}_2$ orbifold mediated by fivebranes.}
\end{center}
\end{figure}

The diagrams in Figure 6 depict a portion of the orbifold 
in which only four of the thirty-two $\a\b$-planes
are shown.  (The entire orbifold would be represented as in Figure 1.)
One should think of these ladder-diagrams as assemblies of the 
individual vertex-diagrams represented in Figure 4.
Thus, vertical lines represent seven-dimensional
$\b$-planes and horizontal lines represent ten-dimensional $\a$-planes.
It is further understood that each of
the two ten-planes supports local $E_8$ matter.  The explicit factors of
$E_8$ shown in the first diagram of Figure 6 indicate that the 
ten-dimensional $E_8$ matter is 
completely unbroken on each of the four $\a\b$ intersections shown in
that diagram.  Similarly, any explicit group shown at
a vertex in a ladder-diagram indicates the subgroup ${\cal G}\subset E_8$
which remains unbroken by $\b$ at that indicated vertex. The 
\includegraphics[width=.15in,angle=0]{x.eps}'s
indicate fivebranes. Each has a worldvolume which fills the 
six noncompact dimensions extending out of the plane of the diagram
and carries unit magnetic charge.  
 
The fivebranes
are free to move about the diagrams.  Notably, due to the consistent vertex
indicated at the top of the right-hand column in Figure 4, the fivebranes
are free to move to, and wrap, any of the vertices in 
the first diagram of Figure 6.  
Such a procedure is shown in the second diagram in Figure 6,
where the arrow indicates that the fivebrane has moved to, 
and wrapped, the indicated vertex.  
Note that, in this process,  the local magnetic charge is 
increased by one unit, from $-3/4$ to $+1/4$. In addition, 
the twisted spectrum at that vertex is augmented by one tensor and 
one hypermultiplet. All of this is consistent
with the absorption of a fivebrane. Transitions of this type can occur,
without further constraints, at any vertex of a global orbifold configuration.

We would now like to consider the case where a fivebrane moves to a vertex and
metamorphizes into a gauge instanton via a small instanton phase transition.
As discussed above, this results in the breaking of $E_{8}$ to one of its
subgroups. Previously in this paper, such transitions have been analyzed at a
single vertex only. However, within the context of a global orbifold
configuration, it is necessary to insure that such a phase transition is
compatible with the structure of the surrounding vertices. This puts
additional, rather strong, constraints on the allowed phase transitions. The
pertinent issue involves the electric charge $\rho$ of the $\beta$-planes. 
Since the unbroken ($E_8$) phases correspond to $\rho=0$ and 
the broken phases ($E_7\times SU(2)$ or $SO(16)$) correspond to
$\rho=1$, and since $\rho$ is associated with an entire
seven-dimensional $\b$-plane (i.e. an entire vertical line in 
one of our ladder-diagrams), it would seem that instanton transitions
on vertices should only occur in pairs. 
According to this interpretation, the lone $n_T$=1 vertex 
shown in the second diagram in Figure 6 can smoothly connect to an
instanton only if another fivebrane first moves 
to the complementary six-plane on the top of the ladder-diagram,
as depicted in the third diagram in Figure 6.
\begin{figure}
\begin{center}
\begin{tabular}{c}
\includegraphics[width=2in,angle=0]{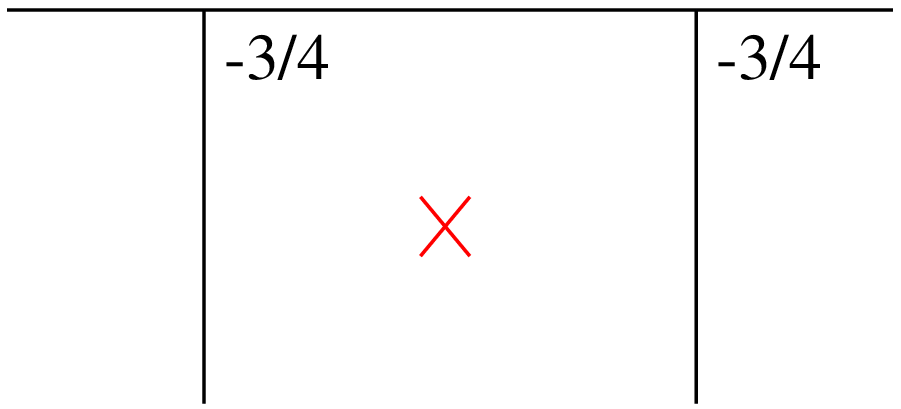} \\[.15in]
\includegraphics[width=2in,angle=0]{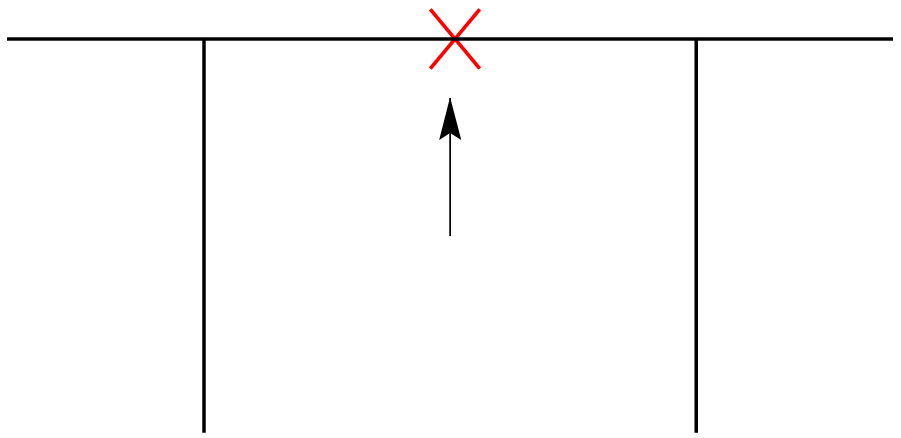} \\[.15in]
\includegraphics[width=2in,angle=0]{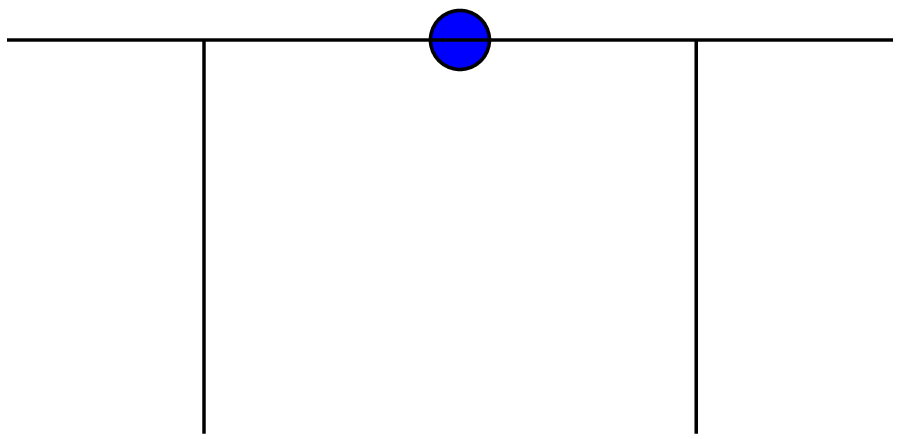} \\[.15in]
\includegraphics[width=2in,angle=0]{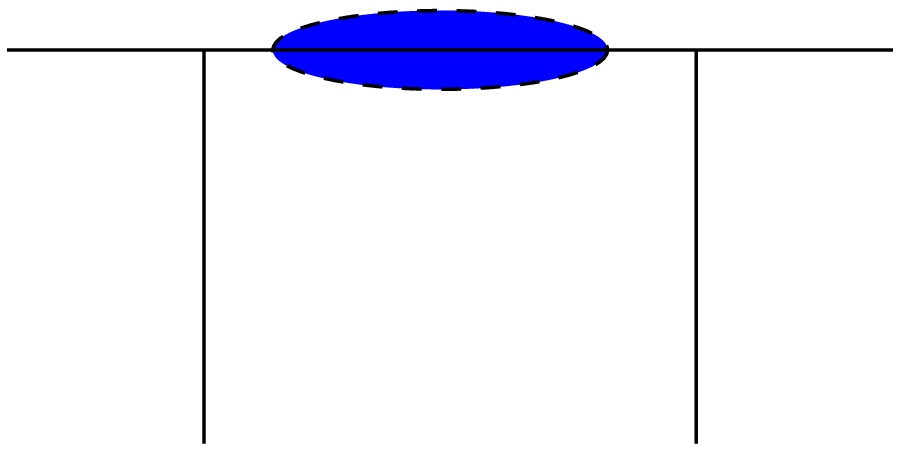} \\[.15in]
\includegraphics[width=2in,angle=0]{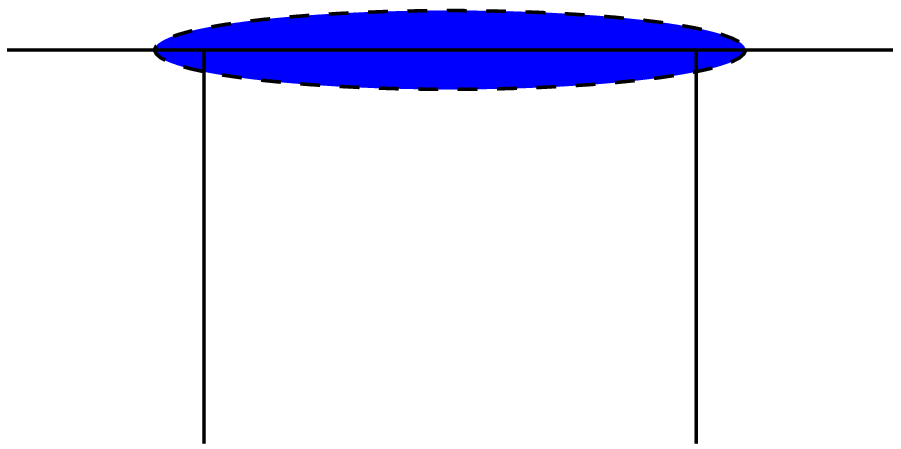} \\[.15in]
\includegraphics[width=2in,angle=0]{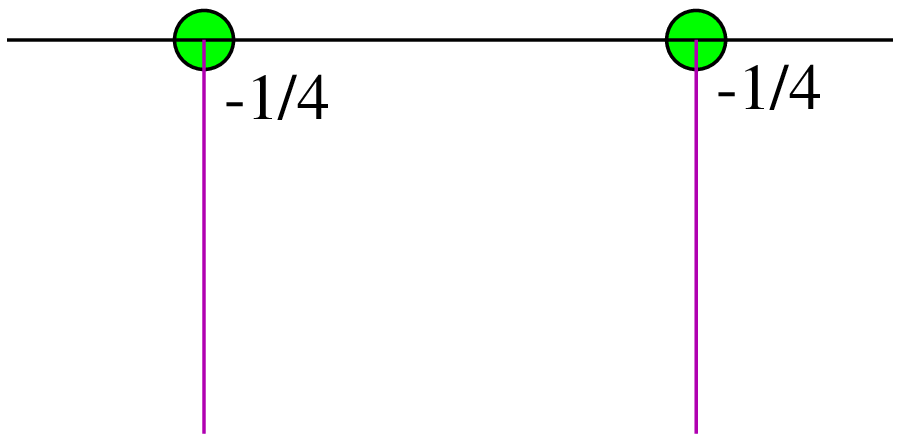} 
\end{tabular}\\[.2in]
\parbox{5in}{Figure 7: A fivebrane spawning a pair of 
half-integrally charged gauge instantons in a process involving
an intermediate integrally charged instanton which grows to
encompass two adjacent $\a\b$-planes.}
\end{center}
\end{figure}
This enables each of the fivebranes to then
smoothly connect to instantons, 
through processes of the sort shown in Figure 6, 
simultaneously turning on an electric charge 
$\rho=1$ on the interpolating seven-plane.  Thus, the sequence
depicted in Figure 5 describes a fivebrane-mediated transition
which smoothly connects an $E_8\times E_8$ 
phase of the moduli space
to an $SO(16)\times SO(16)$ phase.

It is puzzling that the value of the electric parameter $\rho$, 
which is ostensibly derived from a seven-dimensional Chern-Simons 
coupling, changes from
zero to one when the described transitions take place.  This is
puzzling because if $\rho$ is a mere coupling constant, its value
should not be subject to intermittent change.  (On the other hand,
we {\it are} able to justify a related change in the magnetic charge $g$,
since this has an understandable topological origin, which allows us
to resolve such a change in the manner described previously.) 
We suppose that this issue has an interesting resolution. 
This curiosity was independently noticed and commented on
in \cite{ksty}.  For the time
being we allow a situation-dependent $\rho$ as an allowed rule.
We hope to discuss this issue further in a future paper. 
  
It is less clear how the half-integer instantons can emerge via smooth
transitions involving fivebranes.  One picture which suggests itself,
however, is the following.  We can imagine a fivebrane moving to one
of the ten-planes, which smoothly connects in moduli space to a
small ten-dimensional $E_8$ gauge instanton. 
Such a small instanton could then grow until it encompasses each of 
two fixed six-planes within
the ten plane. We could imagine  that this instanton then splits into 
two half-integer instantons, one localized on each of the
two adjacent $\a\b$-planes.  This process is shown by a 
sequence of diagrams in Figure 7. We abbreviate this transition by the 
two-step sequence depicted in Figure 8.
We can describe this process heuristically by 
saying that half of a fivebrane has wrapped
each of the two involved $\a\b$-planes. 
If such a thing were possible, then we could describe another 
phase transition as indicated by the sequence of diagrams in Figure 9.
This depicts a fivebrane-mediated transition which smoothly
connects an $E_8\times E_8$ phase of moduli space to an
$SO(16)\times E_7\times SU(2)$ phase.

\begin{figure}
\begin{center}
\includegraphics[width=2.6in,angle=0]{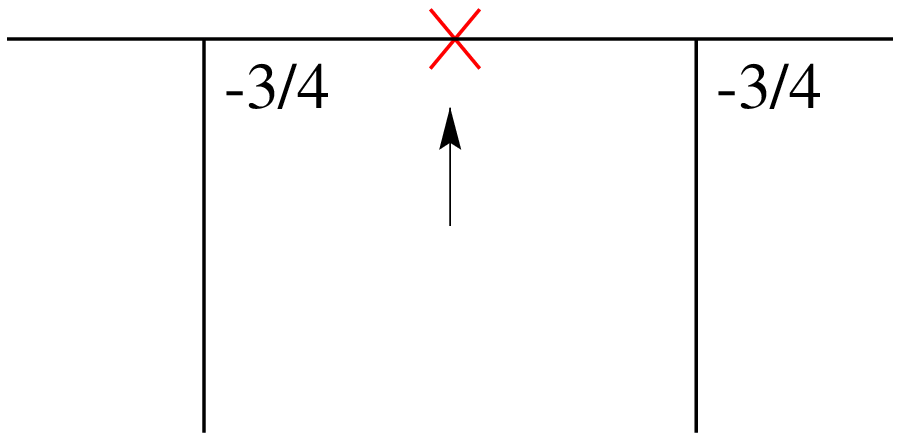} 
\hspace{.3in}
\includegraphics[width=2.6in,angle=0]{other5.eps} \\[.2in]
\parbox{5.3in}{Figure 8: A two-step abbreviation of the process depicted
in Figure 7.  Heuristically, half of a fivebrane is wrapping each of
the indicated vertices.}
\end{center}
\end{figure} 
 
\begin{figure}
\begin{center}
\begin{tabular}{c}
\\[.4in]
\includegraphics[width=2.6in,angle=0]{basic.eps} \\[.3in]
\includegraphics[width=2.6in,angle=0]{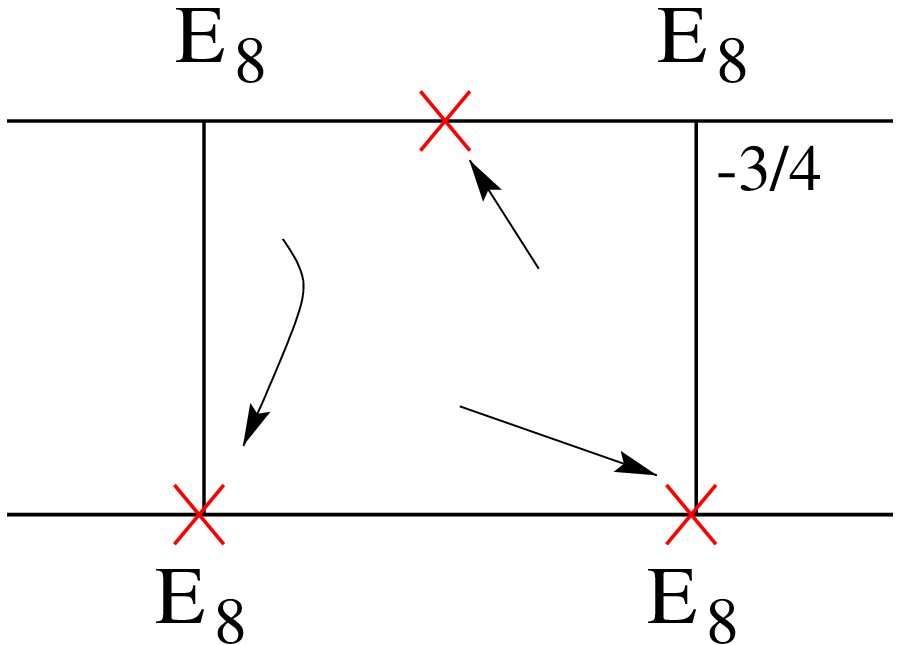} \\[.3in]
\includegraphics[width=2.6in,angle=0]{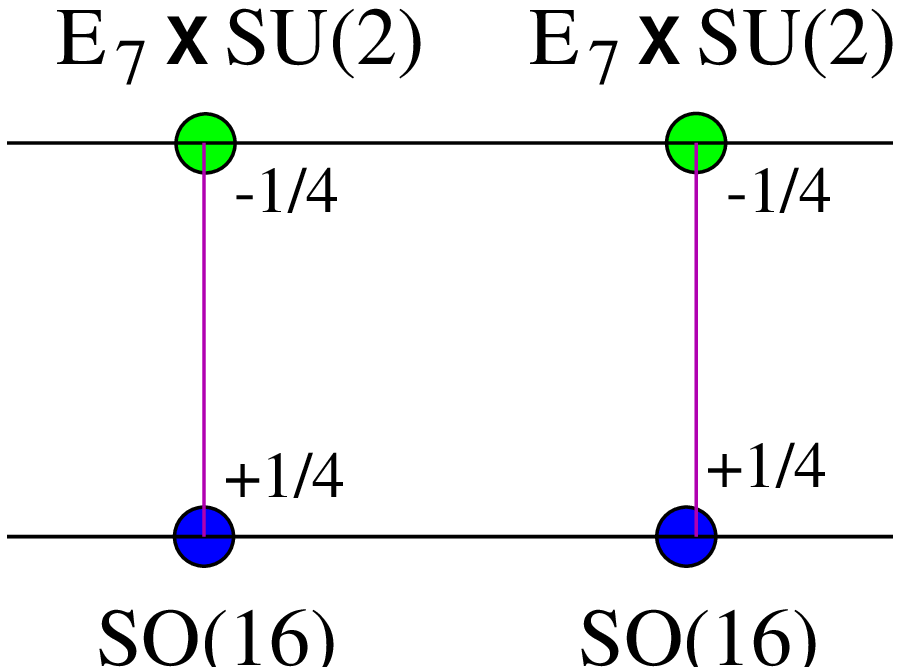} \\[.3in]
\end{tabular}\\[.2in]
\parbox{5in}{Figure 9: An global picture of a phase transition in
the $S^1/{\bf Z}_2\times T^4/{\bf Z}_2$ orbifold, mediated by 
fivebranes.}
\end{center}
\end{figure}
 
\section{Conclusions}
We have described some of the technology needed to resolve 
microscopic consistency issues in {\it M}-theory orbifolds.
We have applied these techniques particularly to the 
$S^1/{\bf Z}_2\times T^4/{\bf Z}_2$ orbifold, and have 
presented an interesting set of consistent vertices, describing 
the twisted states residing on orbifold planes at intersecting points.
By assembling consistent vertices we are able to build up
consistent global orbifolds which describe different phases of 
moduli space.  Our construction rather nicely indicates the
possibility of phase transitions involving {\it M}-fivebranes
as mediators.  There remain intriguing conceptual issues which  
we intend to discuss and hope to resolve in forthcoming papers. 

By resolving a larger set of consistent local vertices, thereby 
enlarging the number of diagrams in Figure 4, we should be 
able to significantly enrich our understanding of the possible
allowed global configurations, and of the implied 
interconnectedness amongst phases in the moduli spaces 
associated with various orbifold compactifications  
indicated by the transitions which our diagrams 
suggest.   It should prove interesting
to apply our techniques to other orbifolds as well. 
\vspace{.3in}
 
\noindent
{\bf Acknowledgements}\\[.1in]
This work is supported by the European Commission TMR programme 
ERBFMRX-CT96-0045, and the work of B.A.O. is supported by the
Alexander von Humboldt-foundation.
We would like to acknowledge helpful emails
from Luis Alvarez-Gaume,
Paul Aspinwall, Bernard de Wit and Bert Schellekens, and 
helpful comments by Vadim Kaplunovsky.  In addition, we are
grateful to Stefan Theisen for pointing out an error in the
previous version of this paper pertaining to the pseudoreality
of a hypermultiplet representation.

\end{document}